\documentclass[11pt,a4paper]{article} 
\usepackage{amsmath}     
\usepackage{amssymb}
\usepackage{amsthm}
\usepackage{eurosym}
\usepackage{setspace}
\usepackage{url}
\usepackage{epigraph}
\usepackage[dvipsnames]{xcolor}

\usepackage{natbib}

\usepackage{color}

\usepackage[text]{diff} 

 \usepackage{graphicx}
\usepackage{caption}
\usepackage{float}
\usepackage{subfig}             
\newcommand\T{\rule{0pt}{2.6ex}}
\newcommand\B{\rule[-1.2ex]{0pt}{0pt}}

       \usepackage[a4paper]{anysize} 
       \marginsize{2.5cm}{2.5cm}{2.5cm}{.8cm}

\usepackage{hyperref}
\hypersetup{
	colorlinks,
	linkcolor={red!50!black},
	citecolor={blue!50!black},
	urlcolor={blue!80!black}
}
\numberwithin{equation}{section}  

\theoremstyle{plain}
\newtheorem{theorem}{Theorem}[section]    

\newtheorem{statement}{Statement}[section]

\theoremstyle{definition}

\theoremstyle{remark}
\newtheorem{remark}{Remark}[section]

\newtheorem{condition}{Condition}[section]

\begin{document}
\linespread{1.45}\small\normalsize

\title{\textsc{\Large{The Precautionary Principle and the Innovation Principle:\\ Incompatible Guides for AI Innovation Governance?%
}}}

 \author{Kim Kaivanto\thanks{e-mail: k.kaivanto@lancaster.ac.uk} \\
 Lancaster University, Lancaster LA1 4YX, UK}

\date{\emph{this version:} \today}

\maketitle

\begin{center}{\bf Abstract}\end{center}  
\begin{spacing}{1.2}

In policy debates concerning the governance and regulation of Artificial Intelligence (AI), both the Precautionary Principle (PP) and the Innovation Principle (IP) are advocated by their respective interest groups. 
Do these principles offer wholly incompatible and contradictory guidance? 
Does one \emph{necessarily} negate the other? 
I argue here that provided attention is restricted to \emph{weak-form} PP and IP, the answer to both of these questions is ``No.''
The essence of these weak formulations is the requirement to fully account for type-I error costs arising from erroneously preventing the innovation's diffusion through society (i.e.\ mistaken regulatory red-lighting) \emph{as well as} the type-II error costs arising from erroneously allowing the innovation to diffuse through society (i.e.\ mistaken regulatory green-lighting). 
Within the Signal Detection Theory (SDT) model developed here, weak-PP red-light (weak-IP green-light) determinations are optimal for sufficiently small (large) ratios of expected type-I to type-II error costs. 
For intermediate expected cost ratios, an amber-light `wait-and-monitor' policy is optimal.  
\emph{Regulatory sandbox} instruments allow AI testing and experimentation to take place within a structured environment of limited duration and societal scale, whereby the expected cost ratio falls within the `wait-and-monitor' range. 
Through sandboxing regulators and innovating firms learn more about the expected cost ratio, and what respective adaptations --- of regulation, of technical solution, of business model, or combination thereof, if any --- are needed to keep the ratio out of the weak-PP red-light zone. 
Nevertheless AI foundation models are ill-suited for regulatory sandboxing as their general-purpose nature precludes credible identification of misclassification costs. 




\vfill
\smallskip\smallskip\smallskip
\noindent \emph{Keywords:}\/\;
artificial intelligence; 
foundational AI;
general-purpose AI systems;
AI governance; 
precautionary principle;
innovation principle; 
countervailing risk; 
scientific uncertainty; 
signal detection theory;
misclassification costs;
discriminability;
ROC curve;
de minimis risk;
trust and polarization;
protected values;
non-comparable values;
continuity axiom;
regulatory sandboxes

\vspace{0.2cm}
\noindent \emph{JEL classification:}\/\;
D81, 
O31, 
O33, 
O38 

\medskip
\end{spacing}
\pagebreak
\linespread{1.45}\small\normalsize


\section{INTRODUCTION}
At one extreme of the popular imagination, unfettered development of Artificial Intelligence (AI) promises previously inconceivable betterment of the human condition. 
At the other extreme, unfettered development of general-purpose AI is associated with extinction-level subjugation of humanity by superintelligence \citep{bostrom:14}.
In-between, entire categories of labor --- both blue-collar (e.g. drivers of taxis, public transport, farm equipment, and road haulage) and white-collar (e.g. accountants, lawyers, and investment advisors) --- are at high risk of being replaced by narrow AI applications. 
There are real questions as to the risks posed by AI to citizens' digital security, physical security (e.g. autonomous weapons), political security, as well as to the ability of society's communication infrastructure to maintain the integrity of information in the face of `deepfakes' \citep{brundage/etal:18}. 
Realization of these risks would in turn undermine the perceived integrity of (`trust' in) society's key institutions \citep{schneier:12}. 
Shannon Vallor \citeyearpar{vallor:16} summarizes that in the 21$^{\textnormal{st}}$ Century, AI will be 
(i) a significant \emph{magnifier} of economic,\footnote{Accentuating inequality in the real economy, but also creating new systemic risks and amplifying existing systemic risks in the financial economy. \citet{danielsson/etal:20} note that AI ``has the potential to destabilise the financial system, creating new tail risks and amplifying existing ones due to procyclicality, endogenous complexity, optimisation against the system and the need to trust the AI engine.''} political, and physical risk, as well as
(ii) an important tool for \emph{managing \& mitigating} these same forms of risk. 

In the face of these risks 
there have been calls to take precautionary measures immediately, rather than wait for full scientific certainty, by which time the window of opportunity for avoiding those bad outcomes will have closed \citep{brundage/etal:18,kuziemski:18,delcastillo:20-fb}.

Responding to these calls for application of the Precautionary Principle (PP), the Information Technology \& Innovation Foundation (ITIF) in the United States has countered by arguing that AI should fall under the Innovation Principle (IP) instead \citep{castro/mclaughlin:19}. 
Meanwhile in the European Union (EU), the European Commission embedded the IP into its procedures for evaluating AI innovation \citep{ec:18},\footnote{``For any new regulatory proposals that shall be needed to address emerging issues resulting from AI and related technologies, the Commission applies the Innovation Principle, a set of tools and guidelines that was developed to ensure that all Commission initiatives are innovation friendly:
[\cite{ec:16} \url{https://data.europa.eu/doi/10.2872/626511}]'' \citep[][p 15, fn 56]{ec:18}.}
and it served as a guiding principle throughout the development of the \emph{EU Artificial Intelligence Act} \citeyearpar{ec:24}.

Do these principles offer wholly incompatible and contradictory guidance for the governance and regulation of AI? 
Does one \emph{necessarily} negate the other? 
Some advocates appear to think so. 
How can government reconcile these seemingly contradictory demands? 

In this paper I argue that among the multitude of formulations --- there are more than 20 definitions of PP alone --- the \emph{weak} forms of PP and IP are compatible. 
This is due to the weak forms' focus on cost structure, where these costs are `comparable' in a decision-theoretic sense, whereas the stronger forms of PP and IP feature `protected values' which are non-comparable. 
Formally, the latter violate the continuity axiom, whereby it is not possible to strike trade-offs between them and other values. 
For this reason, the stronger forms of PP and IP offer conflicting, incompatible guidance for AI governance. 

An important consequence of the focus on type-I and type-II error cost structure is that it also necessitates pinning down explicit candidate interventions --- because the costs of mistaken intervention depend on the specific characteristics of the intervention that is implemented. 
This specificity resolves the open problem of what types of actions can be considered precautionary, and what specific interventions are to be implemented in different PP (or IP) applications \citep{bodansky:04}. 
Context and implementability considerations determine the set of potential interventions, each of which has its own type-I and type-II error cost structure. 
If this set of candidate interventions is non-empty, Benefit-Cost Analysis may be applied across the candidates to select the socially `best' intervention. 

The key novel insight underpinning this paper is that one of the two corner solutions in Signal Detection Theory (SDT) corresponds to non-postponement of weak-PP red-light intervention, while the remaining corner solution corresponds to non-postponement of weak-IP green-lighting. 
The corner solutions occur when the expected error cost structure is highly asymmetric. 
This paper addresses the question of just how asymmetric the cost structure must be in order to trigger weak-PP red-light (or weak-IP green-light) intervention. 
The answer depends on the distinguishing power of current AI-risk science, captured in the discriminability parameter $d^\prime$, as well as an inferential-surprise-tolerance parameter $\epsilon$.\footnote{For details, see Section \ref{sec:non-postp}, p \pageref{tolerance-param}.} 
Superficial resemblance aside, the latter parameter should not be confused with a \emph{de minismis} threshold. 

Careful consideration of SDT mathematical structure yields further insights into the consequences of interactions between base rates, expected error cost structure, and discriminability $d^\prime$. 
For instance when public trust in science is eroded, from the public's standpoint the relevant discriminability parameter is zero. 
If one sets $d^\prime = 0$ in SDT, it follows that for all less-than-unity ($<1$) expected error cost ratios, weak-PP red-light intervention is optimal, whereas for all greater-than-unity ($>1$) expected error cost ratios, weak-IP green-lighting is optimal. 
And if the erosion of public trust extends to the work of risk professionals and regulatory institutions, then each respective interest group will prioritize their own perception of the expected error cost ratio and the associated red-light or green-light determination --- i.e. the result is societal polarization by interest group, even without the effect of non-comparable protected values. 

The SDT-based model can also be used to explain how `regulatory sandboxes' structure the risk-regulation problem to allow weak-PP red-lighting and weak-IP green-lighting as possible outcomes within a single coherent framework. 
As an institutional innovation, sandboxing has been growing in popularity among regulators around the world. 
This regulatory instrument allows AI testing and experimentation to take place within a structured environment of limited duration and societal scale --- crucially, constraining the expected cost ratio --- so that the innovative firm begins its sandbox journey within the SDT-model's amber-light `wait-and-monitor' range. 
And throughout, as the regulator closely monitors testing and experimentation, it is well-placed to make a weak-PP red-light determination or a weak-IP green-light determination. 
By the end of the sandbox journey, regulators and innovating firms learn more about the expected cost ratio, and what respective adaptations --- of regulation, of technical solution, of business model, or combination thereof, if any --- are needed to keep the ratio out of the weak-PP red-light zone. 
Regulatory sandboxes are real-world examples of institutions in which the \emph{weak forms} of PP and IP coexist without contradiction, consistent with the SDT-based model's predictions. 
But the SDT-based model also provides guidance on the limits to implementing weak-form PP and IP with regulatory sandboxes --- which is particularly germane to initiatives that aim to create regulatory sandboxes for general-purpose AI systems, also known as `AI foundation models'. 

In the sequel, Section \ref{sec:principles} reviews the definitions and interpretations of PP and IP.
Section \ref{sec:classical-sdt} introduces SDT-based optimal inferential thresholds for policy decisions. 
Section \ref{sec:non-postp} develops necessary conditions for non-postponement of weak-PP red-lighting and weak-IP green-lighting determinations within SDT. 
Section \ref{sec:comprehensive} clarifies the nature of comprehensive cost estimation required for implementing weak-PP and weak-IP determinations within SDT, as well as the method for solving the associated intervention-selection problem. 
Section \ref{sec:discussion} contains the discussion, and Section \ref{sec:conclusion} concludes.

\section{PRINCIPLES}\label{sec:principles}
In a strict sense it is misleading to refer to \emph{the} PP and \emph{the} IP, as there are many different variants and interpretations of each.

\subsection{Precautionary Principles}
Of the twenty-plus definitions of the PP in existence \citep{sandin:99-heraij,lofstedt/etal:02-JPAM,sunstein:05}, I focus here on three key spinal points in an ascending scale of stringency: weak PP, strong PP, and super-strong PP.

The PP emerged in the former West Germany during the 1970s%
\footnote{Some authors trace the PP's emergence further back, to the late 1960s, in West Germany, Sweden and Switzerland; by the late 1970s the PP was already being cited in U.S. law as well as in international agreements \citep{lofstedt/etal:02-JPAM,wiener/etal:11}.}
as part of the country's social-democratic-planning response to large-scale environmental problems including acid rain, pollution of the North Sea, and climate change \citep{boehmer-christiansen:94,o'riordan/jordan:95-ev,defur/kaszuba:02-ste}. 
Section VII of the Ministerial Declaration announced in London at the conclusion of the 1987 Second International Conference on the Protection of the North Sea included the following statement of the PP:
\begin{quote}
Accepting that, in order to protect the North Sea from possibly damaging effects of the most dangerous substances, a precautionary approach is necessary which may require action to control inputs of such substances even before a causal link has been established by absolutely clear scientific evidence. \citep{ilm-27-835-88}
\end{quote}
But the most widely known variant of the PP was adopted as Principle 15 of the 1992 UNCED Declaration on Environment and Development (the Rio Declaration):
\begin{quote}
In order to protect the environment, the precautionary approach shall be widely applied by States according to their capabilities. Where there are threats of serious or irreversible damage, lack of full scientific certainty shall not be used as a reason for postponing cost-effective measures to prevent environmental degradation. \citep{ilm-31-874-92}
\end{quote}
This is regarded as the definitive articulation of the \emph{weak PP}.
A slightly more verbose restatement of it appears in Article 3 of the United Nations Framework Convention on Climate Change.
Under the weak variant of the PP, there is no mention of which party bears the burden of proof.

\vspace{0.2cm}
\begin{remark}\label{rem:PP-under-risk-too}
Under the weak PP, preventive intervention is undertaken even though ``full scientific certainty'' is lacking.
Although Knightian uncertainty satisfies this condition, it is not synonymous with ``lack of full scientific certainty''.
Knightian-uncertainty-based formalizations are consistent with the weak PP, but not exclusively so.
The weak PP does not exclude the formalization of scientific uncertainty as a unique, if possibly high-dispersion probability distribution.
Hence the weak PP is a candidate for being extended to the domain of risk. 
\end{remark}

\vspace{0.2cm}
\begin{remark}\label{rem:non-postp}
The combination of threats of serious or irreversible damage --- i.e.\ non-trivial (mis-)classification costs --- and scientific uncertainty entails that there is no critical experiment currently available to resolve the societal-level intervention/nonintervention question. 
The weak PP asserts that in these circumstances cost-effective intervention measures shall not be postponed. 
In this sense the weak PP requires \emph{non-postponement} of preventive intervention. 
\end{remark}

\vspace{0.2cm}
\begin{remark}\label{rem:cost-effective-measures}
The weak PP explicitly restricts preventive intervention to ``cost-effective measures''.
Under one interpretation, this refers to positive net-benefit measures established through Benefit-Cost Analysis (BCA).
Under a second interpretation, this refers to an intervention selected using Cost-Effectiveness Analysis (CEA).
The latter CEA method emerged in response to some of the methodological problems as well as the perceived ethical problems involved in assigning monetary valuations to non-market benefits.
However CEA implicitly invokes a separate set of assumptions, including in particular
(i) that Willingness To Pay (WTP) per effectiveness unit is constant for all levels of effectivity and
(ii) that this WTP is identical for all members of the population \citep{johannesson:95-ssm,garber/phelps:97-jhe,bleichrodt/quiggin:99-jhe,dolan/edlin:02-jhe}.
For these reasons, I henceforth interpret references to `cost-effective measures' through the lens of BCA.
\end{remark}

\vspace{0.2cm}
A \emph{strong} PP variant was articulated in the Wingspread Consensus Statement on the Precautionary Principle  (the Wingspread Statement), which was signed by all 32 scientists, philosophers, lawyers and environmental activists who participated in the Science and Environmental Health Network's January 24--26 1998 Conference on the Precautionary Principle held in the Wingspread Conference Center, Racine, WI:
\begin{quote}
When an activity raises threats of harm to human health or the environment, precautionary measures should be taken even if some cause-and-effect relationships are not fully established scientifically. In this context the proponent of an activity, rather than the public, should bear the burden of proof.\footnote{\url{http://www.sehn.org/wing.html}}
\end{quote}
Unlike the weak PP, the strong PP
(i) does not mention costs,
(ii) does not acknowledge that different nation-states have different levels of resources (`capabilities') available for environmental protection, and
(iii) does not limit preventive intervention to threats of serious or irreversible harm.
The strong PP employs the operative word `should', which can refer to either the moral duty (moral imperative) for action, or the moral desirability of action.
Hence from the text of the strong PP alone, it is not clear
(a) whether preventive intervention is called for as a moral, categorical imperative, regardless of the direct and indirect (opportunity) costs of implementing preventive intervention,
or
(b) whether preventive intervention is called for as being desirable, yet subject to the practical direct- and indirect-cost tradeoffs within the totality of obligations involved in running a nation state, given its resources and degree of economic development.
From context we may infer that Wingspread Conference attendees intended the former, moral-imperative interpretation.
But this is not evident from the text of the Wingspread Statement alone.
Finally --- yet crucially --- the strong PP explicitly imposes the burden of proof on the proponent of an activity. 

The \emph{super-strong} PP precludes aforementioned ambiguity by specifying not only the burden of proof, but also the standard of proof:
\begin{quote}
the [PP] mandates that when there is a risk of significant health or environmental damage to others or to future generations, and when there is scientific uncertainty as to the nature of that damage or the likelihood of the risk, then \emph{decisions should be made so as to prevent such activities from being conducted unless and until scientific evidence shows that the damage will not occur.} [emphasis added] \citep{blackwelder:02} 
\end{quote}
Thus under the super-strong PP preventive intervention is the default condition when
(i) there is a risk --- any risk --- of significant harm and
(ii) there is scientific uncertainty over the level or probability of that harm.
The burden of proof lies with those who wish to proceed with the potentially harmful activity.
The standard of proof required by the super-strong PP is extreme, in that preventive intervention remains in place ``until scientific evidence shows that the damage will not occur.''
This exceeds even the highest standard of proof employed in US Common Law, which is `beyond a reasonable doubt', corresponding to a Bayesian posterior probability greater than $99\%$ but less than $100\%$ \citep{weiss:03-lpr,weiss:03-iea,weiss:06-erl}.
Neither is it the `full conviction of the judge' ($90\%$, $95\%$, or $99.8\%$) standard of proof employed in continental European Civil Law \citep{schweizer:15}. 
A literal reading of the super-strong PP requires a $100\%$ standard of proof to be achieved before preventive intervention may be withdrawn.

Henceforth, this paper focuses on the weak-form PP. 

\subsection{Innovation Principles}

Since 2013, representatives of the European Regulation \& Innovation Forum (ERIF)\footnote{\url{https://www.eriforum.eu/}} grouping of industrial concerns in the European Union (EU) have advocated a new risk principle, dubbed the \emph{Innovation Principle}, to complement the PP: 
\begin{quote}
...whenever the EU's institutions consider policy or regulation proposals, the impact on innovation should be fully assessed and addressed \citep{hudig:15-002}.
\end{quote}
Some have viewed the creation and promotion of the IP as a vehicle though which corporate lobbyists are attempting to balance, if not neutralize, application of the PP in the EU \citep{garnett/etal:18-lit,delcastillo:20-fb,ducuing:22}. 
However, a conservative reading of the wording suggests no more than comprehensive, non-myopic cost estimation requiring a full inventory of direct and opportunity costs.  
In particular, the IP requires full accounting of the false-positive cost components associated with forgone diffusion of the innovation through the economy. 
Upon this reading, the IP sits naturally within the weak-PP framework.  

In contrast, the final exhortation to \emph{fully address} the impact on innovation could have  expansive implications under a less conservative reading. 
Conceivably, one set of expectations as to how impacts on innovation would be `fully addressed' involve the prevention of all such negative impacts outright. 
This in turn suggests that underlying valuations of false-positive misclassification costs are so extreme so as to invoke `protected values' and associated infinite costs. 

Thus with the IP, as with the PP, some interpretations spill over from the domain of normative rationality, characterized by comparability and the continuity axiom (\emph{weak IP}) into the domain of protected values, characterized by non-comparability and violation of the continuity axiom (\emph{super-strong IP}). 

An example of the latter super-strong IP is contained in the Information Technology \& Innovation Foundation's (ITIF's) 2019 publication, ``Ten ways the Precautionary Principle undermines progress in Artificial Intelligence'' \citep{castro/mclaughlin:19}.
The authors urge policymakers to embrace the `hope-based' IP, and to thoroughly reject the fear-, doubt-, and anxiety-based PP. 
Although lacking a concise formal definition of the IP, they argue that ``speculative concerns [about potential harms] should not hold back concrete benefits'' of AI innovation. 
Moreover, the authors draw from the work of Adam Thierer, a popularizer of \emph{permissionless innovation}: 
\begin{quote}
...the notion that experimentation with new technologies and business models should generally be permitted by default. Unless a compelling case can be made that a new invention will bring serious harm to society, innovation should be allowed to continue unabated and problems, if any develop, can be addressed later. \citep{thierer:16}   
\end{quote}
In \citeauthor{castro/mclaughlin:19}'s \citeyearpar{castro/mclaughlin:19} ITIF formulation, the IP is a mirror image of super-strong PP, designed to supplant the PP for the governance of AI innovation. 

In the European and World Trade Organization context, \citet{portuese/pillot:18-mjiel} advance economic and legal arguments for the promotion of an IP to counterbalance the prevailing PP and to shift the burden of proof. 
This too therefore is also a super-strong IP variant, but \citet{portuese/pillot:18-mjiel} envision a continual judicial balancing exercise between the IP and the PP, in the interest of serving the ideal of justice.

Even within the context of European Union (EU) legislation and regulation, there is no commonly agreed definition of the IP \citep{renda/simonelli:19, renda/pelkmans:23}. 
P\={e}teris Zilgalvis, a Judge at the General Court of the European Union, characterizes the IP in regulatory impact assessment as ``prioritizing regulatory approaches that serve to promote innovation while also addressing other regulatory aims'' \citeyearpar{zilgalvis:25-nlr}.
Within EU regulatory impact assessment then, the IP is conceived in terms that are not intrinsically incompatible with other regulatory principles such as the PP. 
Meanwhile within the EU's research \& innovation laws and regulations, the IP is defined not in terms of specific criteria for intervention or non-intervention, but in terms of `innovation-friendliness' of legislation and regulation preparation, review, and revision: 
\begin{quote}
The Innovation Principle is a tool to help achieve EU policy objectives by ensuring that legislation is designed in a way that creates the best possible conditions for innovation to flourish.

The principle means that in future when the Commission develops new initiatives it will take into account the effect on innovation.

This will ensure that all new EU policy or regulations support innovation and that the regulatory framework in Europe is innovation-friendly. \citep{ec:19}
\end{quote}
The EU's commitment to innovation-friendliness is currently implemented through 
(i) Research and Innovation Tool  \#21 in the Better Regulation Toolbox, 
(ii) innovation deals, which reveal obstacles to innovation, making it possible for future revision of regulation to address these obstacles, 
(iii) foresight and horizon scanning to anticipate future trends and enable anticipatory policymaking, 
(iv) impact-assessment support, and 
(v) the introduction of regulatory sandboxes in Member States \citep{ec:24} to allow a responsive, experimental approach to regulation for e.g. FinTech, blockchain/distributed ledger technologies, artificial intelligence, and the internet of things. 
In AI-specific policy documents the European Commission (EC) reiterates this innovation-principle derived innovation-friendliness \citep{ec:18}. 
This approach is also being adopted in the United Kingdom, where the Department for Science, Innovation and Technology (DSIT) has produced guidance for AI regulation under the ``pro-innovation principle'' banner \citep{dsit:23,dsit:24}. 

Although a diverse range of activity types are being carried out under the IP rubric, henceforth this paper focuses on the \emph{weak-form IP}, which we conceive in terms that parallel those of weak-form PP: 
(a) weak-IP green-lighting is undertaken even though full scientific certainty is lacking over eventual benefits or harms of the innovation; (b) no critical experiment is currently available to resolve the societal-level intervention/nonintervention question; (c) cost-effectiveness is a constraint for both the weak IP as well as the regulators implementing the weak IP, which we interpret here through the lens of BCA; and (d) silence on which party bears the burden of proof.

\subsubsection{Regulatory sandboxes}\label{sec:reg-sandboxes}
The progenitor of present-day regulatory sandboxes, Project Catalyst, was launched in 2012 by the U.S. Consumer Financial Protection Bureau (CFPB) with the purpose of promoting innovative consumer-friendly financial solutions \citep{cfpb:12}. 
The term was popularised by the U.K. Financial Conduct Authority's (FCA's) Project Innovate, which launched a ``regulatory sandbox'' in 2016 to provide a supportive environment for fintech innovation. 
A 2020 World Bank survey found 73 fintech-focussed sandboxes in operation in 57 countries \citep{wb:20}. 
Four years later the number of active sandboxes worldwide stood at 142 \citep{markellos/etal:24}.\footnote{The Centre for Competition Policy (CCP) Regulatory Sandboxes Database may be filtered by sandbox Status: Announced, Closed, Considering, Open, Suspended, and Unknown. The figure of 142 is obtained by filtering by `Open' status. \url{https://competitionpolicy.ac.uk/research-projects/portal-on-regulatory-sandboxes/}} 
Around the world, regulatory sandboxes cater not only to financial technology (fintech) innovation, but also to distributed ledger innovation, the internet of things innovation, digital authentication and verification innovation, communications innovation, energy innovation, public health innovation, transport innovation, and insurance technology (insurtech) innovation. 
We are not aware of any active regulatory sandboxes catering to pure AI innovation.\footnote{The CCP Regulatory Sandbox Database includes one entry for a 2021 announcement of the possibility of an EU-wide AI Sandbox. As of March 2025, this AI-dedicated sandbox had not been created.} 
The flow of new sandbox initiatives is ongoing. 
For instance Israel launched a new regulatory sandbox program in January 2025 with the stated purpose of boosting innovation, and sandboxing is included as a key component in the country's National AI Program.\footnote{\url{https://aiisrael.org.il/ }, accessed 4 March 2025.} 

According to the World Bank study, a \emph{regulatory sandbox}
\begin{quote}
...is typically a virtual environment that enables 
live testing of new products or services in a 
controlled and time-bound manner [usually 6 months duration]. Controlled 
experimentation in a live environment provides 
a structured approach to promoting innovation 
and guiding interactions with firms while 
allowing regulators good oversight of emerging 
financial products. Regulatory sandboxes are 
open to innovative business models, products, 
and processes, whether regulated, unregulated, 
or slated for possible future regulation. ...[and] 
are usually classified into four types, based on 
their objectives: (i) policy-focused;
(ii) product- or innovation-focused; 
(iii) thematic; and 
(iv) cross-border. \citep{wb:20}
\end{quote}

\noindent Most regulatory sandboxes are created to satisfy multiple objectives. 
Firm-sandbox relationships are collaborative, offering the firms informal `steers' and `regulatory comfort' as part of an individually tailored plan whereby temporary derogations can be granted without altering existing rules outside the sandbox. 
Regulator-operated sandboxes offer the opportunity for regulatory experimentation --- to learn what regulatory solutions are suitable for new technologies and novel business models, while still satisfying the regulator's societal and legal obligations. 
Once sufficient regulatory experimentation and associated experience is gained, operation of the sandbox may be contracted out to a service provider. 
For example the U.K. FCA's permanent digital sandbox is currently operated under contract by fintech company NayaOne Ltd.\footnote{\url{https://nayaone.com/}} 
Alongside the digital sandbox, NayaOne offers complementary services, including synthetic data to enable secure testing and experimentation without risking data privacy \& protection or commercial confidentiality conflicts. 

The OECD's 2023 survey of AI regulatory sandboxes found that the number of regulatory sandbox initiatives around the world had grown to 100, with AI featuring in the technological propositions of many sandbox startups \citep{oecd:23}.
According to the Centre for Competition Policy's Regulatory Sandboxes Database, the total number of sandbox initiatives around the world (past, current, planned) numbers 199, spanning 92 countries  \citep{markellos/etal:24}.

Among the new entries, the U.K. Information Commissioner's Office (ICO) developed a regulatory sandbox in 2021 to support firms launching products and services reliant upon the safe and legal use of personal data, with particular reference to compliance with the U.K. General Data Protection Regulation (UK GDPR) and the U.K. Data Protection Act 2018 (DPA 18). 
AI is a key technological enabler of these firms, though the areas of focus are diverse: 
Central Bank Digital Currencies (CBDCs), commercial use of drones, consumer health technology (healthtech), decentralised finance (DEFI), genomics, immersive technology and virtual worlds, neurotechnologies, next-generation Internet of Things (IoT), next-generation search, personalized AI, and quantum computing. 

And in May 2024, the U.K. Medicines and Healthcare products Regulatory Agency (MHRA) launched `AI Airlock', a new regulatory sandbox to help the Agency to identify and address the challenges of regulating standalone AI as Medical Devices (AIaMDs). 
The initial cohort of projects are virtual or real-world projects, and simulation exercises within the sandbox allow MHRA to test a range of regulatory issues these devices face when deployed within the U.K. National Health Service (NHS). 
AI Airlock therefore is narrowly focused on AIaMDs intended for the U.K. NHS. 

Explicit in-scoping of AI `foundation models'\footnote{
The U.S. draft AI Foundation Model Transparency Act (2023) defines a foundation model as ``an artificial intelligence model trained on broad data, generally uses self-supervision, generally contains at least 1,000,000,000 parameters, is applicable across a wide range of contexts, and exhibits, or could be easily modified to exhibit, high levels of performance at tasks that could pose a serious risk to security, national economic security, national public health or safety, or any combination of those matters.'' 

Generative language models are the most prominent category of foundation models at the time of this writing. 
Google launched its Bidirectional Encoder Representations from Transformers (BERT) Large Language Model (LLM) in October 2018. 
BERT has 340 million parameters and a 16 GB, 3.3 billion word training dataset. 
Five years later, OpenAI's mixture-of-experts architecture GPT-4 features 17.6 trillion parameters trained on 13 trillion tokens ($\approx$ 9.75 trillion words). 
In January 2025 China's DeepSeek released their DeepSeek-R1 model, which has a sparse mixture-of-experts architecture with 671 billion parameters (of which 37 billion active per token), trained with extensive use of large-scale reinforcement learning on a dataset of 14.8 trillion tokens. 1 million tokens $\approx$ 750,000 words; therefore 14.8 trillion tokens $\approx$ 11.1 trillion words. 
}
(also known as `frontier AI systems' or `general-purpose AI systems') is notably absent from the \citet{oecd:23} global survey of sandboxes, the Centre for Competition Policy Regulatory Sandbox Database \citep{markellos/etal:24},\footnote{\url{https://competitionpolicy.ac.uk/research-projects/portal-on-regulatory-sandboxes/}} as well as from the mission statements of the U.K.-based sandboxes. 

The same observation also applies to the European Union's major legislative effort, the EU AI Act \citep{ec:24}.\footnote{The EU AI Act (2024) includes separate chapters (Artiles 53 and 54) on general-purpose AI systems. See \url{https://artificialintelligenceact.eu/article/53/} and \url{https://artificialintelligenceact.eu/article/54/}.}
Article 57 of the EU AI Act requires each Member State (possibly in co-operation with the competent authorities of other Member States) to establish at least one ``AI regulatory sandbox'' at the national level, to be in operation by 2 August 2026. 
Within EU legislation, the introduction of regulatory sandboxes is understood as one of the key means of implementing the IP. 
\begin{quote}
These sandboxes are controlled environments where AI systems can be developed, tested, and validated before being released to the market. The goal is to foster innovation while identifying and mitigating any risks, particularly those related to fundamental
rights, health, and safety. The sandboxes will also provide guidance on regulatory expectations and requirements. If an AI system is successfully tested in a sandbox, the provider can use this as proof of compliance with regulations. The sandboxes are also intended to facilitate cross-border cooperation and share best practices.” \citep[summary of Article 57]{ec:24}
\end{quote}

Nevertheless, much of the fast-moving, high-stakes, large-scale investment in general-purpose AI foundation model innovation is taking place in the U.S. and in China, where there currently are no AI-focussed regulatory sandboxes.\footnote{Eleven countries are listed in \citet{oecd:23} as having AI-related regulatory sandboxes: Canada, Colombia, Estonia, France, Germany, Korea, Lithuania, Malta, Norway, Singapore, and U.K. (Annex B: AI-related sandboxes, pp. 30--32.)}
The \emph{general-purpose} nature of AI foundation models such as LLMs poses a specific challenge for weak-form PP and IP, which we discuss in Section \ref{sec:discussion} below.

A number of legal scholars have produced academic analysis and commentary upon evolving regulatory sandbox policy initiatives in connection with the IP and other aspects of regulation \citep{renda/simonelli:19,ducuing:22,truby/etal:22-ejrr,truby/etal:22-lit,mccarthy:23-lit,yordanova/bertels:24}. 
The present paper aims to complement this legal scholarship with a decision-theoretic model from which both weak-PP and weak-IP arise naturally as particular solutions, and which furthermore identifies the essential features of regulatory sandbox instruments that enable them to implement weak-form PP and IP for \emph{specific applications} of AI foundation models, but not for the AI foundation models themselves.

\section{Optimal inferential thresholds for policy decisions}\label{sec:classical-sdt}
Some of the problems inherent in Null Hypothesis Significance Testing (NHST) as currently practiced\footnote{See Appendix \ref{app:NHST}} can be addressed through incorporation of a context-dependent loss function into the determination of an appropriate $\alpha$ level to be used within the Neyman-Pearson lemma.
Among the numerous approaches to incorporating error costs into statistical inference, the simplest --- and one which has the advantage of being consistent with Neyman and Pearson's frequentist approach --- is known as Signal Detection Theory (SDT) \citep{egan:75,green/swets:66,macmillan/creelman:91}. 
The core elements of SDT, in addition to the above-mentioned ROC curve, are
(i) the misclassification cost matrix,
(ii) the objective function under which the inferential threshold is to be optimized, and
(iii) the population prevalence rates of the conditions captured in $H_0$ and $H_1$ respectively, i.e.\ the parameters in frequentist statistics which correspond to Bayesian prior probabilities for $H_0$ and $H_1$.

I begin by introducing the confusion matrix, entries of which consist of True Positives (TP), False Negatives (FN), False Positives (FP) and True Negatives (TN) counts obtained from repeated application of a specific threshold $x^\prime$ (see Table \ref{tab:conf-matrix}).
It is common to re-express these entries as row-specific (within-hypothesis) rates:
$TPR=TP/(TP+FN)$, $FNR=FN/(TP+FN)$, $FPR=FP/(FP+TN)$, $TNR=TN/(FP+TN)$.
Associated with each cell of the confusion matrix is a corresponding misclassification cost, which is independent of the value of the threshold $x^\prime$ employed to generate the confusion matrix (see Table \ref{tab:cost-matrix}).
The essence of `context' is represented via a particular set of misclassification costs.
For the purpose of presenting SDT, misclassification costs are assumed to be measured or estimated in an unbiased manner, reflecting overall societal concerns.
This entails unbiased accounting for both immediate (i.e.\ generation-specific) as well as intertemporal (i.e.\ inter-generational) externalities.

\begin{table}[!ht]
\caption{Classification matrices.}
\centering
\begin{tabular}{cc}
 \subfloat[1][Confusion matrix (counts).]{
\label{tab:conf-matrix}
\small\normalsize\small
\begin{tabular}{cc|c|c|}
                              &            & \multicolumn{2}{c}{Inference under $x^\prime$} \\
                              &            &   $\lnot H_0$ &   $H_0$  \T\B\\
\hline
\raisebox{-1.9ex}[0pt][0pt]{Actual}&$H_1$  & ~~~$TP$~~~  & ~~~$FN$~~~ \T\B\\
\cline{2-4}
                              &   $H_0$    &    $FP$     &    $TN$ \T\B\\
\cline{2-4}
\end{tabular}
\normalsize\small\normalsize
}
&
 \subfloat[2][Misclassification cost matrix.]{
\label{tab:cost-matrix}
\small\normalsize\small
\begin{tabular}{cc|c|c|}
                              &            & \multicolumn{2}{c}{Inference} \\
                              &            &   $\lnot H_0$ &   $H_0$  \T\B\\
\hline
\raisebox{-1.9ex}[0pt][0pt]{Actual}&$H_1$  & ~~~$C_{TP}$~~~  & ~~~$C_{FN}$~~~ \T\B\\
\cline{2-4}
                              &   $H_0$    &    $C_{FP}$     &    $C_{TN}$ \T\B\\
\cline{2-4}
\end{tabular}
\normalsize\small\normalsize
}
\end{tabular}
\end{table}

Letting $N$ denote the total number of observations in the (random) sample $TP+FN+FP+TN=N$, then the sample-based estimates of the population prevalence rates may be written as\; $P(H_0)=(FP+TN)/N$\; and\; $P(H_1)=(TP+FN)/N$.

With few exceptions \citep{kaivanto:14-ra}, applications of SDT are couched in terms of \emph{minimizing expected misclassification cost}.
The central results of classical SDT are all derived under this `expected misclassification cost' objective function.
For present purposes the parsimony and tractability of this objective function serve well. 
Adopting an expected utility objective function introduces additional structure within the square brackets of \eqref{eq:OOP-isocost-slope}, but  the rest of the qualitative structure remains intact. 

The optimally chosen cutoff threshold\; $x^*$\; minimizes expected misclassification costs  $E(C)$ subject to the constrained relationship between the TPR and the FPR, which may be represented with the twice-differentiable function\; $G:[0,1]\to[0,1]$.
This function, written as\; $\mathrm{TPR} = G(\mathrm{FPR})$, captures the ROC curve.
As $N$ grows larger,  $\lim_{N\to\infty}TPR= 1-\beta$ and $\lim_{N\to\infty}FPR=\alpha$, which in turn are defined by
\begin{align}
\alpha =&\; P(X>x^\prime\,|\,\theta_0) =\; \int_{x^{\prime}}^{+\infty}f(x|\theta_0)\,\textnormal{d}x
 \label{eq:alpha-defined}\\
1-\beta=&\; P(X>x^\prime\,|\,\theta_1) =\; \int_{x^{\prime}}^{+\infty}f(x|\theta_1)\,\textnormal{d}x\;\;. \label{eq:power-defined}
\end{align}
The slope at a point on the ROC curve determined parametrically by $x^\prime$ is given by the derivative at the point $x^\prime$
\begin{equation}\label{eq:likelihood-ratio}
\diff*{P(X>x^\prime\,|\,\theta_1)}{P(X>x^\prime\,|\,\theta_0)}{x^\prime}\;= \;\frac{-f(x^\prime|\theta_1)}{-f(x^\prime|\theta_0)}\;=\;l(x^\prime)\;\;,
\end{equation}
which is the likelihood ratio at\; $x^\prime$.
I assume $G^\prime>0$ and $G^{\prime\prime}<0$, ensuring that the monotone-likelihood ratio condition holds.\footnote{Note that $G^{\prime\prime}<0$ is not satisfied by arbitrary combinations of sampling distributions. When both distributions are Gaussian, $G^{\prime\prime}<0$ is satisfied everywhere in the support of $x$ only when the two sampling distributions have the same variance \citep{hills/berbaum:11-ar}. }

\begin{figure}[!ht]
\caption{ROC curve, assuming $\sigma_0=\sigma_1=1$ and four different discriminability parameters.}
\label{fig:ROC-and-slopes}
\centering
\begin{tabular}{c}
   \scalebox{.4}[.4]{\includegraphics{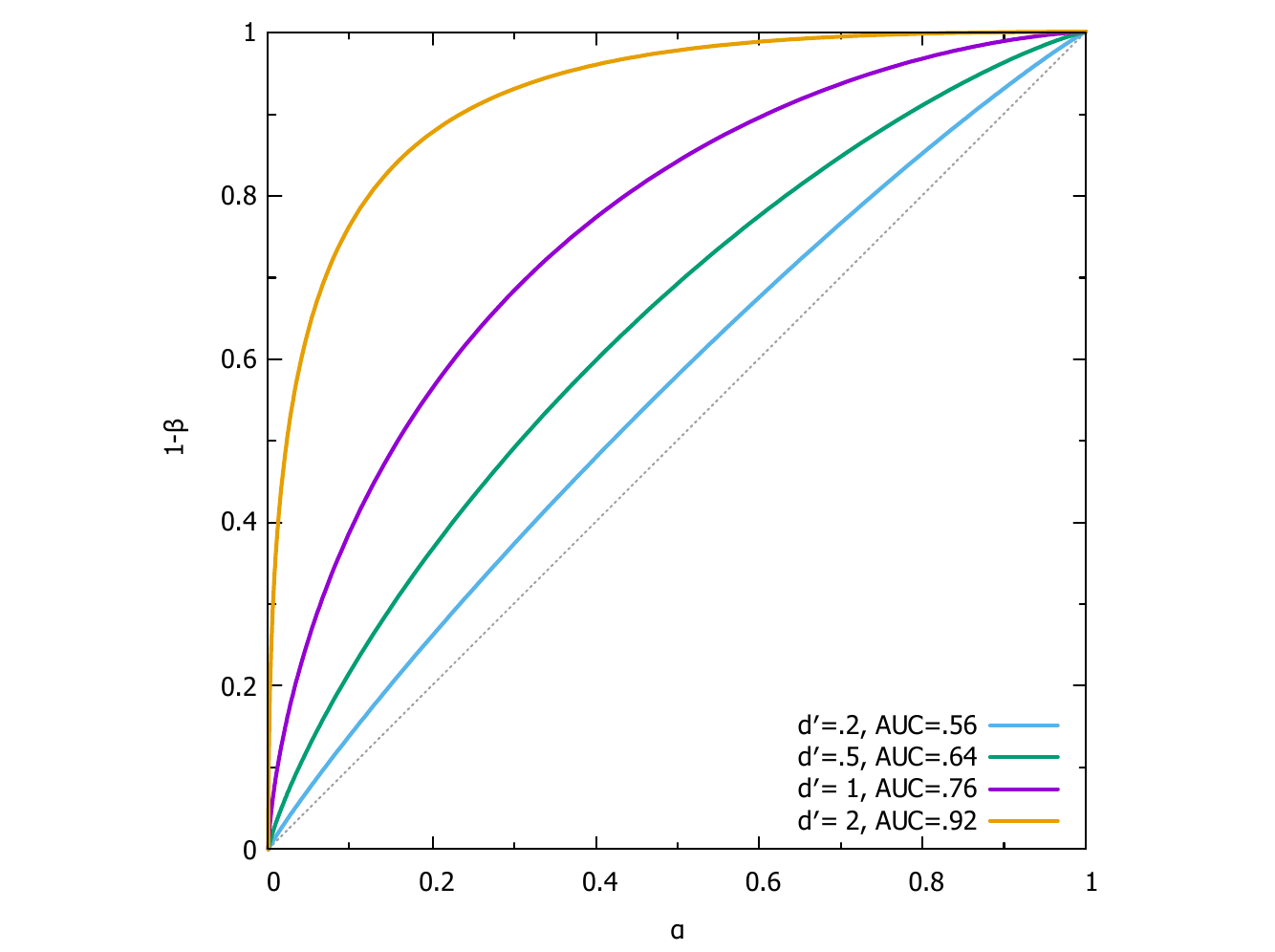}}
 \label{fig:ROC-curves}
\end{tabular}
\end{figure}

Solving the constrained-minimization problem
\begin{equation}\label{eq:optim}
\min_{x^\prime}E(C)\qquad\textnormal{s.t.}\qquad 1-\beta = G(\alpha)
\end{equation}
gives the optimality condition
\begin{equation}\label{eq:OOP-isocost-slope}
l(x^*) =
\frac{P(\theta_0)}{P(\theta_1)}
\left[\frac{C_{\mathrm{FP}}-C_{\mathrm{TN}}}{C_{\mathrm{FN}}-C_{\mathrm{TP}}}\right] =
\diff*{\mathrm{(1-\beta)}}{\mathrm{\alpha}}{\bar{C}^*}\;\;,
\end{equation}
which states that the slope of the cost-minimizing iso-expect-cost line at the optimal operating point is given by the ratio of the expected opportunity cost of misclassifying a Negative to the expected opportunity cost of misclassifying a Positive.
From \eqref{eq:likelihood-ratio} and \eqref{eq:OOP-isocost-slope} it is also clear that the optimality condition defines the critical likelihood ratio\; $l(x^*)$, and that \eqref{eq:OOP-isocost-slope} is a tangency condition between the least-cost iso-expected-cost line and the ROC curve.
From \eqref{eq:alpha-defined} and \eqref{eq:power-defined} we have that
\begin{align}
\alpha^* =&\; \int_{x^*}^{+\infty}f(x|\theta_0)\,\textnormal{d}x
 \label{eq:optimal-alpha}\\
1-\beta^*=&\;  \int_{x^*}^{+\infty}f(x|\theta_1)\,\textnormal{d}x\;\;. \label{eq:optimal-power}
\end{align}
When the cutoff threshold is optimally determined by \eqref{eq:OOP-isocost-slope}, the associated optimal level of the test $\alpha^*$  responds to changes in misclassification costs and population prevalence rates $P(\theta_0)$ and $P(\theta_1)$.
Setting $\theta_0=0$ WLOG and furthermore assuming Gaussian sampling distributions $X \sim N(\theta_i,1),\; i=\{0,1\}$, $\theta_1 > \theta_0$, the optimal cutoff threshold $x^*$ responds to the remaining parameters as follows:
\begin{equation}\label{eq:optimal-cutoff}
x^*\; =\; \frac{1}{\theta_1}\left( \ln(C_{FP}-C_{TN}) - \ln(C_{FN}-C_{TP})+\ln(P(\theta_0)) - \ln(P(\theta_1)) + \frac{\theta_1^2}{2}\right)\;\;.
\end{equation}
If misclassification costs are symmetrical in the sense that $C_{FP}-C_{TN}=C_{FN}-C_{TP}$ and the base-rate probabilities are also symmetrical $P(\theta_0)=P(\theta_1)$, then the optimal cutoff threshold $x^*$ falls half-way between $\theta_0$ and $\theta_1$, where the two pdfs intersect $f(x^*|\theta_0)=f(x^*|\theta_1)$.
The associated optimal operating point $(\alpha^*, 1-\beta^*)$ is that ROC-curve point that coincides with the minor diagonal, where the slope of the iso-expected-value line is unity $l(x^*)=1$.
Due to the concavity of $\ln(\cdot)$, increasing misclassification-cost increments have a diminishing impact upon $x^*$.
However, the natural logarithm's concavity and limiting value $\lim_{P\to 0^+} \ln(P)=-\infty$ entail that the hypothesis with the smaller base rate has a disproportionately larger impact upon the location of the optimal cutoff threshold.
This responsiveness characteristic of $x^*$, $\alpha^*$ and $(1-\beta^*)$ under SDT sits in contradistinction to their fixed nature under the Neyman-Pearson lemma, i.e.\ $1-\beta^{\text{NP}} = G(0.05)$.

Whereas the $\alpha$ level is arbitrary under NHST, it is optimally adapted to base-rates and misclassification costs in SDT.
Whereas in NHST, the distinction made between $p$-values 0.051 and 0.049 is artificially sharp, under SDT the distinction made between $p$-values $\alpha^*+0.01$ and  $\alpha^*-0.01$ is not artificial, but anchored in real-world consequences ($C_{FP}$, $C_{TN}$, $C_{FN}$, $C_{TP}$) and base rates ($P(\theta_0)$, $P(\theta_1)$).
Finally, whereas statistical significance in NHST is not synonymous with scientific or decision-making significance, rejecting the null hypothesis under SDT's optimal $\alpha^*$ level is, by design, synonymous with decision-making significance.

I conclude this section by noting that the approach embodied in SDT is consistent with David Cox's general exhortations concerning the use of $p$-values.
\begin{quote}
The $P$-value has, \emph{before action or overall conclusion can be reached}, to be combined with any external evidence available and, \emph{in the case of decision-making}, with \emph{assessments of the consequences of various actions} \citep[emphasis added,][]{cox:82-bjcp}. 
\end{quote}

\section{Non-postponement}\label{sec:non-postp}
The weak PP admonishes \emph{against waiting} for a critical experiment before introducing protective intervention when there are threats of serious or irreversible damage and science currently does not offer a definitive, certain answer (see Remark \ref{rem:non-postp} on p. \pageref{rem:non-postp}).
When the current scientific state of the art lacks conclusive resolving power and is accompanied by credible threats of serious damage, scientific uncertainty in itself does not constitute sufficient grounds to suspend or postpone preventive intervention.
In SDT, prevailing scientific uncertainty may be operationalized through sampling distributions under the null and alternative hypotheses. 
Let $H_0:\theta_0=\theta$ be the status-quo level of the critical index variable, and define $H_1:\theta_1=\theta$ ($\theta_1 > \theta_0$) to be the (irreversible) higher value of the critical index-variable induced by a commercial innovation in the state of the world where it is harmful. 
Maintaining the assumption of equal-variance sampling distributions,\footnote{This assumption is done purely for mathematical convenience. The existing SDT literature, now well established, explores the multitude of deviations from this assumption, involving different combinations of underlying distribution family and unequal variance.} the resolving power of state-of-the-art scientific experiments may be summarized succinctly with the \emph{discriminability} index:
\begin{equation}
d^\prime\;=\; \frac{\theta_1 - \theta_0}{\sigma}\;\;.
\end{equation}
A given $\theta_1-\theta_0$ difference can be consequentially large or consequentially small, depending on the value of $\sigma$.
Small absolute effect sizes  $\theta_1-\theta_0$  and large standard deviations --- whether due to limited precision of scientific measurement or due to explicit gaming of the research process by non-independent researchers\footnote{fully developed treatment of which is deferred to future work} --- are associated with small Area Under the Curve, $\textnormal{AUC}=\Phi\left( \frac{d^\prime}{\sqrt{2}}\right)$, where $\Phi$ is the standard normal CDF.
Along the principal diagonal of the ROC space, where $d^\prime$=0 and AUC=0.5, the SDT-based inference performs no no better than chance, as achieved e.g. with the toss of a fair coin.
Larger $d^\prime$ and AUC permit improvement over mere chance (see Figure \ref{fig:ROC-curves}).

In this model, \emph{non-postponement} of the red-light/green-light regulatory response arises when either
\begin{equation}\label{eq:x-neg-infty}
 x^* \to -\infty\qquad\qquad \textnormal{and thus}\qquad\qquad \alpha^*=1\:,\;1-\beta^*=1
\end{equation}
or 
\begin{equation}\label{eq:x-pos-infty}
 x^* \to +\infty\qquad\qquad \textnormal{and thus}\qquad\qquad \alpha^*=0\:,\;1-\beta^*=0
\end{equation}
Under \eqref{eq:x-neg-infty}, any finite $x$ draw falls to the right of $x^*$, and $H_0$ is rejected in favor of $H_1$, implying weak-PP red-light intervention. 
Hence, current experiments will generate scores satisfying $x^* < x$ with probability $p(x^* < x) = 1$.
Similarly, under \eqref{eq:x-pos-infty}, any finite $x$ draw falls to the left of $x^*$, and $H_0$ is not rejected, implying weak-IP green-lighting. 
Hence, current experiments will generate scores satisfying $x < x^*$ with probability $p(x< x^*) = 1$.

In ROC space, non-postponement occurs when either the (1,1) top-right or (0,0) bottom-left corner solution obtains. 
Depending on the combination of discriminability $d^\prime$, prior probability $P(\theta_0), P(\theta_1)$, and misclassification costs $C_{FP}-C_{TN}$, $C_{FN}-C_{TP}$, it is possible for the optimal solution to be a top-right corner solution (non-postponement of weak-PP red-light preventive intervention despite prevailing scientific uncertainty) or a bottom-left corner solution (non-postponement of weak-IP green-lighting, allowing commercial activity to continue despite scientific uncertainty). 

In order for \eqref{eq:x-neg-infty} to hold for an arbitrary $d^\prime > 0$, $\textnormal{AUC} > 0.5$ ROC curve, \emph{either} the numerator of the optimality condition expression \eqref{eq:OOP-isocost-slope} is equal to zero, \emph{or} the denominator is equal to infinity, \emph{or} both. 
In other words the iso-expected cost line must be horizontal. 
In turn for \eqref{eq:x-pos-infty} to hold for an arbitrary $d^\prime > 0$, $\textnormal{AUC} > 0.5$ ROC curve, the numerator of \eqref{eq:OOP-isocost-slope} must be infinite so that the iso-expected cost line is vertical. 

Moral imperatives and trade-off-resistant  `protected values'\footnote{Protected Values (PVs) are characterized by an absolute resistance to trade-offs: they are in this sense `protected' from being subject to trade-offs with other values or attributes. This means that no amount of compensating benefit will induce an individual to make even a small sacrifice to her PV. To illustrate: for an
individual who views ecosystem life as sacrosanct (i.e.\ a PV), there is no finite amount of compensating economic gain that could justify the extinction of a single species. \citep{baron/spranca:97-obhdp}} --- such as those implicit in the strong- and super-strong-form PPs --- can cause individuals to impute infinite values to either $C_{FP}$ or $C_{FN}$. 
However in order to remain consistent with normative decision theory, misclassification costs must remain \emph{finite}. 
For this reason, the slope of the iso-expected-cost line \eqref{eq:OOP-isocost-slope} can be neither infinite nor zero, contrary to the possibilities identified in the preceding paragraph. 

With finite misclassification costs it is nevertheless possible for the optimal solution to be arbitrarily close to the corner solution.  
This can be formalized by setting an arbitrarily small $\epsilon_1\; (\epsilon_1\in\mathbb{R}_{++})$\label{tolerance-param} such that $x^* < x$ with probability tolerance $1-p(x^* < x) < \epsilon_1$ for weak-PP intervention in the $\delta_1$-neighborhood of the top-right corner of the ROC space. 
Similarly, $x < x^*$ with probability tolerance $1-p(x < x^*) < \epsilon_0 \in\mathbb{R}_{++}$ for non-intervention in the $\delta_0$-neighborhood of the bottom-left corner of the ROC space, where $\epsilon_0$ is arbitrarily small. 

\begin{condition}{($\epsilon_1$-tolerance non-postponement)}\label{con:epsilon1-tolerance}
\begin{equation*}
    (\alpha^*, 1-\beta^*)\in B_{\delta_1}(1,1) 
    \qquad \Rightarrow \qquad
    1 - p(x^*<x) < \epsilon_1\;\;,
    \qquad \textnormal{where}\qquad \delta_1= \sqrt{2}\cdot\epsilon_1
\end{equation*}
\end{condition}

This states that when the optimal operating point $(\alpha^*,1-\beta^*)$ falls within the $\delta_1$-neighborhood of the top-right (1,1) corner solution, the probability of any current experiment generating a score value to the right of $x^*$ is within $\epsilon_1$ tolerance of 1. 

\begin{figure}[!ht]
\caption{Iso-expected cost line (in {\color{red}red}) that is tangent with the ROC curve (in {\color{cyan}cyan}) within the $\delta_1=0.004$ neighborhood of (1,1) (in {\color{Plum}magenta}).}
\label{fig:delta_1-neighb}
\centering
\begin{tabular}{c}
   \scalebox{.6}[.6]{\includegraphics{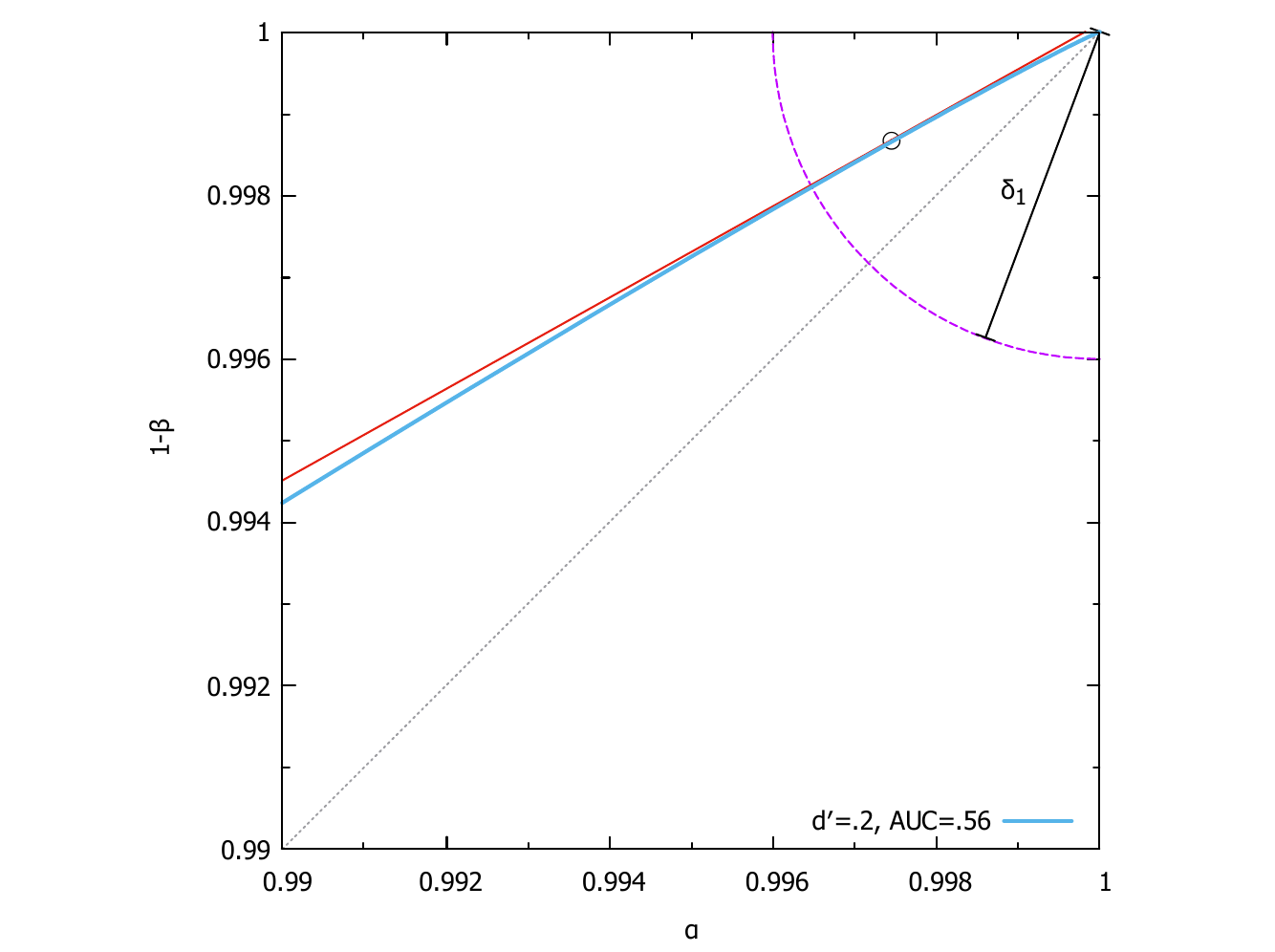}} 
\end{tabular}
\end{figure}

Condition \ref{con:epsilon1-tolerance} is illustrated in Figure \ref{fig:delta_1-neighb}. 
In this figure we let $\delta_1=0.004$ be the proximity threshold representing `arbitrary closeness' to the corner solution. 
The expected incremental costs of type-II error (the denominator in \eqref{eq:OOP-isocost-slope}) are sufficiently large relative to the expected incremental costs of type-I error (the numerator in \eqref{eq:OOP-isocost-slope}), that the iso-expected cost line's point of tangency with the ROC curve falls within the $\delta_1$ neighborhood of the (1,1) corner solution.
\newpage

\begin{condition}{($\epsilon_0$-tolerance non-postponement)}\label{con:epsilon0-tolerance}
\begin{equation*}
    (\alpha^*, 1-\beta^*)\in B_{\delta_0}(0,0) 
    \qquad \Rightarrow \qquad
    1 - p(x<x^*) < \epsilon_0\;\;,
    \qquad \textnormal{where}\qquad \delta_0= \sqrt{2}\cdot\epsilon_0
\end{equation*}
\end{condition}
This states that when the optimal operating point $(\alpha^*,1-\beta^*)$ falls within the $\delta_0$-neighborhood of the bottom-left (0,0) corner solution, the probability of any current experiment generating a score value to the left of $x^*$ is within $\epsilon_0$ tolerance of 1.

\begin{figure}[!ht]
\caption{Iso-expected cost line (in {\color{red}red}) that is tangent with the ROC curve (in {\color{cyan}cyan}) within the $\delta_0=0.004$ neighborhood of (0,0) (in {\color{Plum}magenta}).}
\label{fig:delta_0-neighb}
\centering
\begin{tabular}{c}
   \scalebox{.6}[.6]{\includegraphics{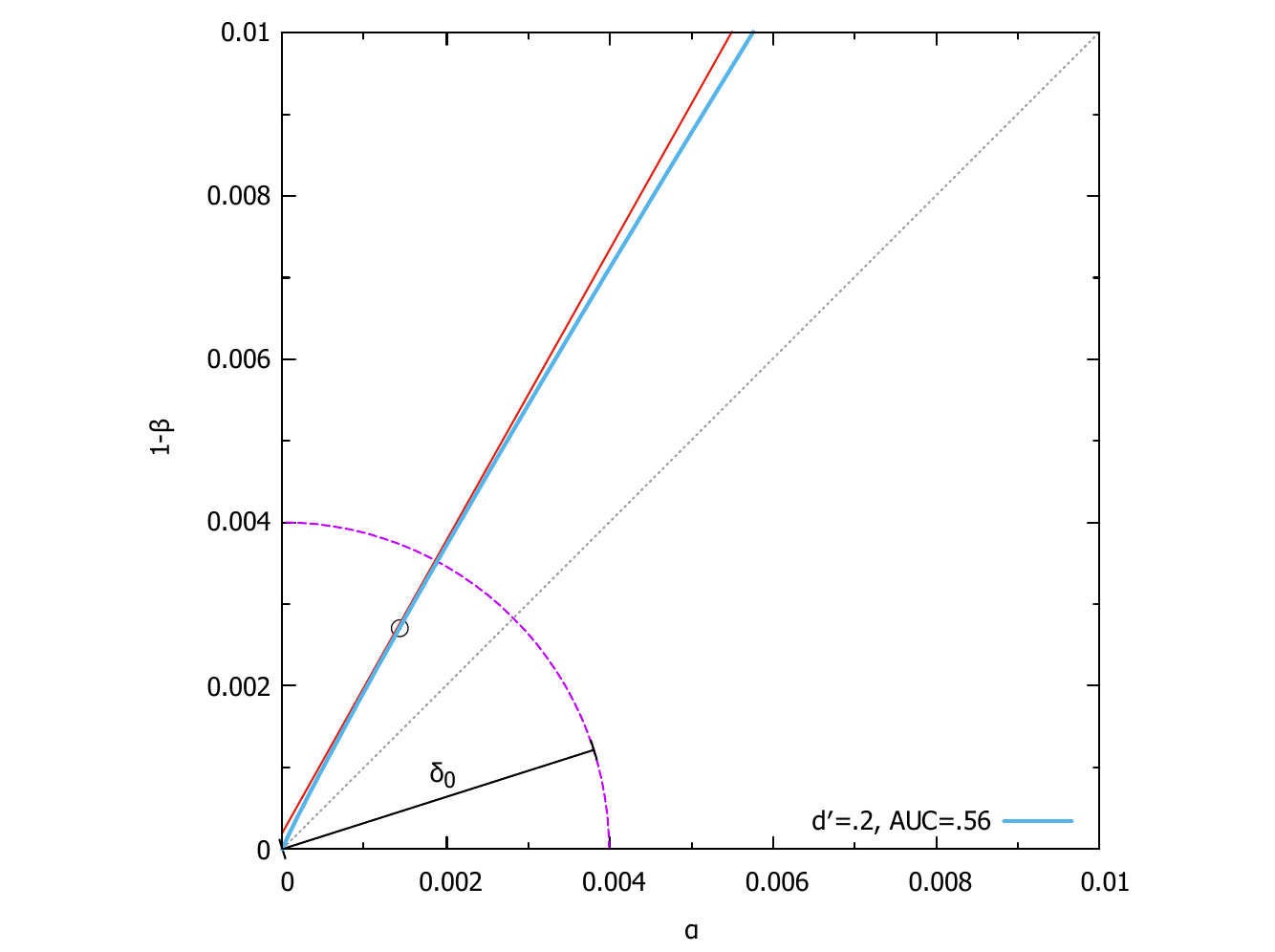}} 
\end{tabular}
\end{figure}

Condition \ref{con:epsilon0-tolerance} is illustrated in Figure \ref{fig:delta_0-neighb}. 
Here $\delta_0=0.004$ is the proximity threshold representing `arbitrary closeness' to the corner solution.
The expected incremental costs of type-II error (the denominator in \eqref{eq:OOP-isocost-slope}) are sufficiently small relative to the expected incremental costs of type-I error (the numerator in \eqref{eq:OOP-isocost-slope}), that the iso-expected cost line's point of tangency with the ROC curve falls within the $\delta_0$ neighborhood of the (0,0) corner solution. 

\newpage

Condition \ref{con:epsilon1-tolerance} formalizes a necessary condition for $\epsilon_1$-tolerance weak-PP red-light preventive intervention. 
In turn, Condition \ref{con:epsilon0-tolerance} formalizes a necessary condition for $\epsilon_0$-tolerance weak-IP green-lighting, despite prevailing scientific uncertainty over whether the commercial activity is harmful or not. 
This second possibility, which has not been modeled decision-theoretically before, arises naturally from the structure of SDT. 

Derivations supporting Conditions \ref{con:epsilon1-tolerance}  and \ref{con:epsilon0-tolerance} are contained in the Appendices \ref{sec:appendix-epsilon1} and \ref{sec:appendix-epsilon0}.

\section{Comprehensive cost estimation and intervention selection}\label{sec:comprehensive}
The misclassification-cost matrix was succinctly introduced in Section \ref{sec:classical-sdt}.
Yet the validity of any particular application of the SDT model hinges critically upon the care and attention with which entries in the misclassification-cost matrix are constructed and estimated.
Comprehensive, non-myopic cost estimation requires full inventory of direct and opportunity costs.
These costs are specific to the intervention that is to be implemented in the event that the null hypothesis is rejected.
Hence the very first step is to fix a specific intervention to be implemented in the event that the null hypothesis is rejected.
As I show below in Section \ref{sec:multiple-interventions}, this does not preclude evaluating multiple candidate interventions or indeed combinations of interventions, and avoids the simplifying assumptions involved in positing stylized continuous-function representations.\footnote{which for tractability reasons must abstract from representing the full complexity of nonlinear and interaction effects}

Sections \ref{sec:scope-of-comprehensive}--\ref{sec:multiple-interventions} below analyze weak-PP red-lighting interventions $s_j$ where $j\in(1,2,...,J)$ and whether the associated solutions fall within the $\delta_1$-neighborhood of the (1,1) corner solution. 
Section \ref{sec:intervention-selection} then presents the Benefit-Cost Analysis needed to select an optimal intervention from among those satisfying the $\delta_1$-neighborhood condition. 
The same logic, adapted to the $\delta_0$-neighborhood of the (0,0) corner solution, can be applied to analyzing weak-IP green-lighting interventions $s_k$ where $k\in(1,2,...,K)$.

\subsection{The scope of `comprehensive'}\label{sec:scope-of-comprehensive}
Misclassification costs may be inventoried at different entity levels:
for different classes of individuals,
for different organizations or regions,
for different state-level and supra-state-level political entities,
or, for society as a whole.
The higher entity levels aggregate across lower-level heterogeneity.
Successive levels thereby resolve heterogeneity, culminating in societal-level analysis, which for present purposes is assumed to yield a well-defined, unique misclassification-cost matrix.

Although even comprehensive cost inventory cannot anticipate unforeseen side effects, it must nevertheless capture direct costs as well as opportunity costs, many of which are not fully deterministic and therefore require separate estimation.
These include not only the ancillary impacts, but also the `countervailing' or `competing' risks associated with the intervention \citep{lave:81,graham/wiener:95,cross:96,viscusi:98,graham:01-jrr}. 
The internationally agreed phaseout of chlorofluorocarbons (CFCs) to prevent further depletion of the ozone layer provides an interesting illustration of such countervailing risks.\footnote{This example draws on \citet{graham/wiener:95}.}
This phaseout was accomplished by substituting CFCs with hydrochlorofluorocarbons (HCFCs) and hydrofluorocarbons (HFCs), which pose respectively one-tenth and zero ozone-depletion potency compared to CFCs.
Yet these compounds pose in turn other problems:
(i) some have been found to be toxic in laboratory-animal testing,
(ii) they can be many times more expensive than CFCs, slowing the diffusion of refrigeration and its health benefits in poorer countries, and
(iii) they may substitute global-warming risk for ozone-depletion risk.
From the standpoint of global warming, the effect of CFCs is roughly neutral:
their global-warming effect via direct radiative forcing is roughly offset by their global-cooling effect via depletion of lower-stratosphere ozone.
In turn the direct global-warming potential (GWP) of HCFCs and HFCs  is approximately one-fifth that of CFCs --- but without an offsetting global-cooling effect.

In the example above, the GWP of CFCs must be reflected in $C_{\text{TN}}$, i.e.\ the state of nature in which the ozone layer is not being depleted by CFCs and CFCs are correctly not phased out, as well as in $C_{\text{FN}}$, the state of nature in which the ozone layer is being depleted by CFCs and CFCs are erroneously not phased out.
In turn the GWP of HCFCs and HFCs must be reflected in the costs $C_{\text{TP}}$ and $C_{\text{FP}}$ where CFCs are correctly, and respectively incorrectly, phased out.
Thus the SDT-based model implements an incremental-risk calculation within each state of nature (within the numerator and within the denominator in square brackets of \eqref{eq:OOP-isocost-slope}) as well as a risk-risk tradeoff between states of nature (through the ratio between the numerator and the denominator in square brackets of \eqref{eq:OOP-isocost-slope}).

\subsection{Option value}
Option value forms an important class of opportunity cost that is an essential component of comprehensive misclassification-cost estimation.
Note that it is possible for the SDT-based model to support the weak-PP corner solution even in regions of the parameter space where option value is zero --- for instance where no improvements in the resolving power of scientific experiments are expected in the near future.
Hence non-zero option value is not a necessary prerequisite for normatively underpinned weak-PP red-light intervention.
Yet where option value is strictly positive, its inclusion into the misclassification-cost matrix is necessary for normative validity of the SDT-based model.

Already \citet{arrow/fisher:74-qje} examined the problem of when to allow the irreversible development of wilderness land when society learns over time, resolving uncertainty about the benefits of preservation. 
This combination of irreversibility, uncertainty resolution (learning), and sequential decision-making structure gives rise to quasi-option value that raises the opportunity cost of development, reflecting the value of information.
Optimal decisions early in the timeline favor flexibility preservation, as later on it will be possible to undertake the irreversible development if the benefits of preservation are revealed to be low.

In turn \citet{gollier/etal:00-jpe} examine the broader sequential problem in which today's actions can irreversibly increase society's future risk, and propose an economic interpretation of the PP rooted in the standard Bayesian framework. 
They show that in this broader framework, irreversibility and learning combine to yield a quasi-option, as a consequence of which more scientific uncertainty optimally induces society to favor more conservative measures (e.g. tighter restrictions on emissions) early in the timeline to allow more flexibility later on.
\citep{gollier/etal:00-jpe} also study additional effects that occur in the irreversible-accumulation model of harmful substances associated with consumption. 
They find that restrictions on the shape of the utility function --- absolute prudence must be at least twice the absolute risk aversion --- ensure that the optimal response to more scientific uncertainty is to restrict current consumption associated with the accumulation of harmful substances.
Alteratively, restrictions on the set of distributions of risk ensure the same policy implication.

Whereas the SDT-based model supports the weak PP interpreted as a (1,1) corner solution within a one-date model, it is also possible to introduce value-of-information quasi-option value into the misclassification cost matrix by explicitly considering the possibility of scientific progress occurring across two or more dates.
This does not require abandoning the risk-neutrality assumption.
Such quasi-option value enters as an opportunity-cost component in $C_{\text{FN}}$.
It is a measure of the value of information: the benefit obtained by making the decision at time $t=1$ with discriminability $d^\prime_1$, rather than at time $t=0$ with discriminability $d^\prime_0$, where $d^\prime_1 \geq d^\prime_0$ due to improved resolving power of scientific experiments. This value of information is foregone, i.e.\ the quasi-option is extinguished, when $H_0$ is falsely accepted, and the harmful commercial activity is erroneously permitted.

\subsection{Multiple potential policy interventions}\label{sec:multiple-interventions}
In this section the preventive intervention under consideration has been `cessation of the harmful commercial activity'.
However this represents only one extreme of a range of possible policy interventions.
More generally, the SDT-based model of weak PP can be used to partition a collection of policy interventions into those that satisfy the $\delta_1$-neighborhood corner-solution condition $(\alpha^*, 1-\beta^*)\in B_{\delta_1}(1,1)$ and those that do not.
To fix notation, consider a particular policy intervention $s_j$ from among a collection of discrete policy interventions\; $S=\{s_1,s_2,...,s_J\}$, where each policy intervention $s_j$ is assigned an index $j$ according to its misclassification-cost ratio:
\begin{equation}\label{ineq:ordering}
\left[\frac{C^j_{\text{FP}}-C^j_{\text{TN}}}{ C^j_{\text{FN}}-C^j_{\text{TP}} }\right]
\geq \left[\frac{C^{j+1}_{\text{FP}}-C^{j+1}_{\text{TN}}}{ C^{j+1}_{\text{FN}}-C^{j+1}_{\text{TP}} }\right]\;\;. \end{equation}
Thus larger values of the index $j$ are associated with smaller misclassification-cost ratios, ceteris paribus.
The collection\; $S$\; may include multiple candidate `maximum-permissible emissions limits', for example.
However, due to non-linear effects and complex interaction effects with rest of the society and the environment, it is not possible to simply assume that the misclassification-cost ratio  $\left[\frac{C_{\text{FP}}-C_{\text{TN}}}{ C_{\text{FN}}-C_{\text{TP}} }\right]$ is monotonic in the permitted emission limit. 
For AI, the equivalent to an emission limit would be the extent of permitted diffusion across jurisdictions and applications. 

With elements of\; $S$\; indexed according to \eqref{ineq:ordering}, any policy intervention\; $s_j$\; that is found to satisfy the $\delta_1$-neighborhood corner-solution condition $(\alpha^*_{s_j}, 1-\beta^*_{s_j})\in B_{\delta_1}(1,1)$ implies that those policy interventions with equally small or smaller misclassification-cost ratios --- i.e.\ $\{s_{j+1},s_{j+2},...,s_J\}$ --- also satisfy the $\delta_1$-neighborhood corner-solution condition, and equally are feasible candidates for weak-PP red-lighting. 
Of course it is also possible to identify the critical policy intervention\; $s_{j^\prime}$
\begin{equation}\label{eq:inclusion-test}
j^\prime\;\; =\;\; \arg\max_{j\in(1,2,...,J)}\;\sqrt{(1-\alpha^*_{s_j})^2 + (\beta^*_{s_j})^2} \qquad\textnormal{s.t.}\qquad (\alpha^*_{s_j}, 1-\beta^*_{s_j})\; \in\; B_{\delta_1}(1,1)
\end{equation}
to the right of which all remaining policy interventions are feasible candidates for weak-PP red-lighting.

However if policy-intervention \emph{combinations} are contemplated, the possibility that the misclassification costs of a combination are not merely additive but non-linearly impounded (e.g. through interaction effects) necessitates
(i) re-estimation of combination-specific misclassification costs, requiring
(ii) re-evaluation through \eqref{eq:OOP-isocost-slope} whether the \emph{combination} satisfies the $\delta_1$-neighborhood corner-solution condition. 
In general, policy makers may wish to consider the power set (excluding the null set) of simple policy interventions $\mathcal{S}=2^S\setminus\{\varnothing\}$,  comprising the union of simple $S$ and combination policy interventions $\mathcal{S}\setminus S$.
The elements of this set $\mathcal{S}$ must be re-indexed according to \eqref{ineq:ordering}, yielding $\mathcal{S} =\{s_1, s_2, ... , s_{|\mathcal{S}|} \}$, among the elements of which the critical policy intervention $s_{j^\prime}\in\mathcal{S}$ may be found by applying \eqref{eq:inclusion-test} over the expanded index list $j\in(1,2,...,|\mathcal{S}|)$.

\subsection{Intervention selection}\label{sec:intervention-selection}
Expansion of the consideration set beyond the singleton introduces a further selection problem:
which policy measure, or indeed which policy-measure combination, should be implemented?
Recall that the weak PP states that \emph{lack of full scientific certainty shall not be used as a reason for postponing cost-effective measures to prevent environmental degradation} \citep{ilm-31-874-92}. 
Recall also that the direct cost of implementing a policy intervention does not enter into the optimality condition \eqref{eq:OOP-isocost-slope} which determines the optimal cutoff \eqref{eq:optimal-cutoff}.
Consequently it is natural to maximize net social benefit across the candidate PP policy interventions $\mathcal{S}_{j\geq j^\prime}$ subject to a non-negativity constraint.
In other words, Benefit-Cost Analysis (BCA) is applied to each candidate PP policy intervention $s_j \in \mathcal{S}_{j\geq j^\prime}$, and society selects for implementation that policy (combination) with the highest strictly positive net social benefit.

Indeed if $\mathcal{S}_{j\geq j^\prime}$ is a singleton, then in order for it to be implemented under the present normative conception of the weak PP, then it also must pass the  net-social-benefit test.
The necessity of this BCA test is encoded in the verbal statement of the weak PP (see above).
Moreover, it is also necessitated by the method of \emph{inframarginal analysis}, which has been developed for problems involving corner solutions, in which the final step is to evaluate candidate corner and interior solutions by BCA \citep{cheng/yang:04-jebo}. 

Sections \ref{sec:multiple-interventions} and \ref{sec:intervention-selection} show that the SDT-based normative model of the weak PP supports multiple candidate policy interventions, different degrees of stringency in permissible exposures or emissions limits, as well as policy-intervention combinations.
Furthermore, the SDT-based normative model naturally culminates in a policy-selection stage that examines the full portfolio of costs and impacts associated with each candidate policy intervention (and all combinations thereof).
Much of the PP literature neglects this final stage, which is necessary for completing a normative analysis of preventive intervention.

\section{DISCUSSION}\label{sec:discussion}
\subsection{Implications for the PP}
Within the present framework it is not meaningful to separate the question, `When is PP-based preventive intervention warranted?', from `What is the preventive intervention being contemplated?'. 
The former can only be answered for specific candidates of the latter.
Different interventions will have different impacts, both direct and ancillary, which result in different misclassification costs and different benefit-cost ratios.
Much of the PP literature has remained silent on what types of actions can be considered precautionary and what specific action(s) are to be deployed under specific applications of the PP \citep{bodansky:04}. 
This paper's contribution hinges on recognizing that by focusing on one specific intervention at a time, misclassification costs can be pinned down to allow evaluation of whether the 
$\delta_1$-neighborhood (1,1) corner solution holds, the $\delta_0$-neighborhood (0,0) corner solution holds,
or neither holds.
The normative analysis culminates with total benefit-cost comparison among those specific policy interventions --- and combinations thereof --- satisfying the relevant corner-solution condition.
Hence the present SDT-based framework provides an integrated formalization not only of the conditions supporting the existence of the weak PP (weak IP), but also of the associated procedure for identifying the societally optimal policy intervention.
In this sense, it offers a solution for the open problem highlighted by \citet{bodansky:04}, among others.

\citet{graham:01-jrr} observes that existing PP-definition variants leave unanswered how a public decision maker should resolve the dilemma that arises when ``a precautionary action might prevent one hazard but induce[s] another hazard.'' 
The present SDT-based operationalization illustrates how such countervailing risks associated with a particular action $s_j$ may be incorporated into the misclassification-cost matrix (see Section \ref{sec:scope-of-comprehensive}).
The SDT-based model of the weak PP offers the following answer to John Graham's question: the hazard introduced by the precautionary action $s_j$ is incorporated into $C_{\text{TP}}$ and $C_{\text{FP}}$.
Since increasing $C_{\text{TP}}$ decreases the denominator of the term in square brackets appearing in expression \eqref{eq:OOP-isocost-slope}, while increasing $C_{\text{FP}}$ increases the numerator, the hazard induced by the precautionary action $s_j$
unambiguously reduces the likelihood of a $\delta_1$-neighborhood corner solution occurring, 
ceteris paribus.
Whether the $(\alpha^*, 1-\beta^*)\in B_{\delta_1}(1,1)$ condition holds following incorporation of the countervailing risk depends on
(i) the initial (pre-incorporation) proximity of the optimal operating point to the boundary of the $\delta_1$-radius neighborhood of (1,1), 
(ii) the local curvature of the ROC curve, and 
(iii) the magnitude of the countervailing risk induced by $s_j$.
Comprehensive cost estimation, inclusive of countervailing risks, diminishes potential bias against innovation. 
With balanced and comprehensive cost estimation, the SDT-formalization of weak PP-IP reconciles promotion and control of technological innovation, which is a characterization of PP advocated by \citet{todt/lujan:14-ra}.


Just because an industrial innovation poses a credible risk of serious damage to the environment, or an AI innovation poses a credible risk to current technical and economic structures, it does not necessarily follow that the optimal operating point will fall within the $\delta_1$-neighborhood of (1,1). 
It is entirely possible for there to be serious risks of damage, yet when expected incremental misclassification costs are computed according to equation \eqref{eq:OOP-isocost-slope}, the resulting optimal operating point remains within the amber-light `wait-and-monitor' interior of the ROC space. 
For smaller discriminabilities and AUCs, however, the likelihood of the optimal operating point reaching the neighbourhood of a corner solution increases, ceteris paribus. 
In the limit, as $d^\prime\to 0$ and AUC$\to 0.5$, even minute asymmetry between the expected costs of false positives and false negatives will lead to a corner solution. 
Indeed when societal trust in the ability of science to deliver credible, politically even-handed positive discriminability $d^\prime > 0$ is eroded, then even under the SDT-based model one expects to see polarization. 
Part of society will have steeper expected misclassification cost ratios, a (0,0) corner solution, and will argue for the removal of all restrictions on AI innovation. 
Another part of society will have flatter expected misclassification cost ratios, a (1,1) corner solution, and will argue for the shuttering of AI innovation. 

As for Boyer-Kassem's \citeyearpar{boyer-kassem:19-epe} question on the general scope of the PP and its relation to other decision rules, the present work's analytical model offers rich potential for conceptual contextualization of the weak PP. 
Consider the \emph{Argument from Inductive Risk} (AIR), long studied in the philosophy of science \citep{rudner:53-pos,douglas:00-pos,douglas:09}.%
\footnote{According to the AIR, the standard of evidence required to accept or reject a hypothesis should be responsive to the \emph{non-epistemic costs} associated with inferential error.}
The AIR's relevance to the PP is well understood among philosophers \citep{steel:14}.
But whereas the AIR literature has focused on whether scientists should (or should not) adjust their inferential thresholds in light of non-epistemic values, only recently has this been operationalized using SDT to show precisely how such adjustments may be made in light of not only misclassification costs, but also prior probabilities \citep{kaivanto/steel:19-pos}. 
The present paper proceeds from the observation that in widespread scientific practice, inferential thresholds are not in fact adjusted to reflect non-epistemic values. 
Instead, the inferential thresholds used in scientific practice --- for many disciplines, $\alpha=0.05$ --- are a long-standing convention reflecting \emph{epistemic values}. 
This leaves a gap --- in suitability and relevance --- between the inferences generated by science, which largely ignore non-epistemic values, and the needs of policy making, which are driven by the need to strike trade-offs between non-epistemic values. 
Hence, one can characterize the PP as a \emph{patch}\footnote{In computer science the term \emph{patch} refers to retrospectively installed update code that repairs, improves or adapts the functioning of an existing piece of software. Previous versions of this paper emphasized this aspect of the PP.} 
for the mismatch between 
(i) the scientific community's fixed-threshold inferential practices and 
(ii) policy makers' need to strike trade-offs among non-epistemic values. 

This \emph{PP-as-a-patch} characterization also yields insight into an even deeper reason for the existence of PPs, as well as insight into PPs' relations to other decision frameworks. 
The inferential conventions and practices of science currently reflect frequentist NHST. 
If instead all scientists and policy makers were full-fledged Bayesians, there would be no need for PPs. 
Since each experiment and empirical study would yield a posterior probability distribution rather than a dichotomous significant/non-significant result, there would be no role for a patch. 
Both scientists and policy makers would use the same posterior probability distribution, but each community would combine it with a different loss function, capturing epistemic and non-epistemic values respectively, in implementing Bayesian Decision Theory. 

It has become widely accepted that reasonable formulations of the PP must be compatible with the \emph{de minimis} risk principle \citep{carter/peterson:15-e}, according to which sufficiently improbable risks, falling under some threshold, should be ignored. 
Yet when examining the \emph{de minimis} risk principle, 
several fundamental problems have been identified \citep{peterson:02-rm,carter/peterson:15-e,lundgren/stefansson:20-ra}. 
The structure of the SDT-based model of weak PP shows that, strictly speaking, it is erroneous to examine base rates in isolation and to apply a threshold to these probabilities. 
Instead, in SDT these probabilities must be combined multiplicatively with their respective incremental misclassification costs and related to the trade-off between $\alpha$ and $(1-\beta)$ encoded in the ROC curve. 
However, as the discussion of equation \eqref{eq:optimal-cutoff} in Section \ref{sec:classical-sdt} points out, base-rate probabilities are unusually powerful `levers' acting upon the location of the \emph{optimal cutoff threshold} $x^*$. 
If either base rate is sufficiently small, its effect will dominate those of finite, non-extreme misclassification costs, and the optimal cutoff threshold will move far into the tail that accommodates the small-probability hypothesis. 
Using the SDT-based model one can derive a \emph{de minimis-like} base-rate-risk threshold that corresponds to the regulator's $\epsilon_1$ parameter, which captures its tolerance to inferential surprise in issuing a weak-PP red-light determination. 
However, this threshold is not unconditional; it also depends upon error costs, the discriminability $d^\prime$ of current risk science, as well as the regulator's above-mentioned inferential-surprise tolerance $\epsilon_1$. 
As such, this would have to be viewed as a \emph{conditional de minimis} risk principle. 

\subsection{Implications for the IP}
The IP, being much younger than the PP, is sometimes viewed as lacking the deeper foundations and legitimacy claimed for the PP. 
There has been less time --- and perhaps, for varied reasons, less interest among scholars --- to develop deeper foundations for the IP. 
This paper argues that within the SDT-based model, the weak IP is theoretically no less well-founded than the weak PP. 
Indeed weak-form PP and IP emerge naturally and simultaneously from the SDT framework. 
Delineation of decision-theoretic (SDT-based) foundations for weak-form IP and its compatibility with weak-form PP is a central contribution of this paper. 

This SDT-based model underpinning weak-form IP is novel within the context of existing IP scholarship, which adduces a variety of motives and arguments for the IP. 
Given the diffuse nature of those motives, some policy makers interpret the IP equally diffusely as a `pro-innovation principle' which requires regulation to be `innovation friendly'. 
The value of the SDT-based model is not so much in providing exact, parametric estimates or a specific formula that regulators can apply slavishly, but in providing a structure within which to reason through questions pertaining to risk regulation and regulator intervention. 
The first distinction that the SDT-based model highlights is between strong-form PP and IP, which are expressions of protected, non-comparable values, and weak-form PP and IP in which costs are finite and comparable. 
Thus, to advocate the weak IP does not entail adopting the position that all AI innovation should be allowed to proceed without hindrance. 
Instead, the SDT model identifies the essential considerations in making a weak-form PP or IP determination: expected misclassification-error costs, the discriminability that captures the distinguishing power of current AI-risk science, and the regulator's tolerance to inferential surprise. 

A case in point is how the SDT model's structure aids in reasoning through the challenges faced in AI innovation governance. 
Should AI innovation governance be guided by the PP, or by the IP? 
Is it coherent for AI innovation governance to be guided by both? 
Furthermore, what types of policies and instruments does this imply for AI innovation governance? 

Answers to the first two questions are clear by now: \emph{weak-form} PP and IP are compatible, implying no incoherence in a regulator guided by both principles. 
As for the third question, policies and instruments can target (i) discriminability, (ii) the misclassification cost ratio, and (iii) the qualitative  (green-light, amber-light, red-light) properties of the regulatory solution. 
Policy makers in most countries now recognize the importance of innovation in increasing national productivity, which is the most important driver of welfare in the long run.\footnote{In Paul Krugman's oft-repeated words, ``Productivity isn't everything, but, in the long run, it is almost everything. A country’s ability to improve its standard of living over time depends almost entirely on its ability to raise its output per worker.'' \citep{krugman:94} }
Accordingly policy makers are not agnostic to the qualitative properties of the regulatory solution for AI innovation governance:  
they prefer to green-light where possible, else to wait-and-monitor, and only to red-light when it is risk-optimal to do so. 
Moreover, optimal classification into green-light and red-light categories is improved (expected misclassification costs are smaller) the larger the discriminability $d^\prime$ --- the increase of which can be enhanced by learning more about the technological underpinnings and business models of the new AI-based products and the firms bringing them to market. 
The same learning process also allows regulators to estimate more precisely the type-I and type-II error cost structures of those technologies and business models. 
Furthermore, by placing temporary bounds on the societal scale at which the AI innovation is deployed --- e.g. virtual deployment, or beta testing on a limited scale, or limited in-sandbox duration, or conditional, time-limited post-sandbox regulatory approval --- the extremity of the expected misclassification-cost ratio is limited, keeping the optimal solution in the wait-and-monitor range of the ROC curve. 
The aforementioned elements all come together in \emph{regulatory sandbox} instruments. 
In this sense the SDT model provides a decision-theoretic foundation for regulatory sandbox instruments.\footnote{Even though in practice regulatory sandbox implementations often bundle additional objectives beyond (i)--(iii), such as facilitating early-stage access to finance \citep{cornelli/etal:24-rof}.}

As noted in Section \ref{sec:reg-sandboxes}, IP-inspired innovation-friendly `AI regulatory sandboxes' are beginning to propagate across countries. 
Where the sandbox operator can identify the application or set of applications that the AI-based innovation will be employed in, the techniques discussed in Section \ref{sec:comprehensive} can be brought to bear to identify the set of possible type-I and type-II error costs with which to solve the SDT model for red-light, amber-light, or green-light classification by the regulator. 
Indeed the collaborative sandbox process allows both the regulator and firms to discover these potential costs, and how these costs vary with different measures that the regulator (or sandbox operator) can take, as well as with different design, implementation, and business-operation measures that the firm can take. 

The regulator and the firm have less control over \emph{how} the innovation will be used in the market. 
At least three different categories of users must be considered: na\"{i}ve, well-behaved, and nefarious. 
Na\"{i}ve users are liable to mis-use the product through oversight unless hard guardrails are built into the product. 
Nefarious third parties may compromise na\"{i}ve users by exploiting their poor cyber hygiene or susceptibility to social engineering. 
Well-behaved users conform with product developers' instructions, while maintaining good cyber hygiene and rebuffing social-engineering attacks. 
Finally, nefarious users may `game' the product itself, steal data revealed by the product, or use the product in ways not intended by the developer. 

AI-based innovations in particular pose new security, regulatory, and societal risks, insofar as they offer so much more scope for unintended use by na\"{i}ve users, nefarious users, and na\"{i}ve users' accounts hijacked by nefarious third parties \citep{anderljung/etal:23}. 
Difficult regulatory questions also arise over the extent to which firms and regulators must anticipate possible \emph{combinations} of AI tools to achieve nefarious effects. 
These questions bring forth profound ethical and legal debates underpinning regulation and its operational implementation. 

Finally, the SDT model also offers guidance on the limits of regulatory sandbox instruments.\label{para:sandbox-limitations-from-SDT} 
The defining characteristic of `AI foundation models' is that they are general-purpose AI systems capable of generating deep-fake text, audio, static images, and video. 
These capabilities are now being harnessed to create AI assistants that can perform numerous conditionally linked operations, including making and responding to telephone calls, as well as  searching and executing tasks on the internet --- with the potential of reaching any internet-connected system. 
Moreover, AI foundation models can be run either online or locally --- so AI foundation models propagate easily across national borders. 
Currently, regulatory sandbox initiatives remain silent on whether AI foundation models fall within their scope or not (see Section \ref{sec:reg-sandboxes}). 
To the extent that regulatory sandboxes are explained by the SDT model, an attempt to bring AI foundation models within the scope of regulatory sandboxing would constitute over-reach. 
Given the general-purpose nature of AI foundation models, the set of possible applications and the range of legitimate and nefarious uses are vast and open-ended. 
It is therefore not possible to explicitly identify all possible interventions and the associated misclassification costs required in the SDT-based approach. 
Hence if regulatory sandboxing is viable for AI foundation models, it must be based on a different theoretical underpinning than the SDT model developed here. 

\subsection{Contestability}
Finally, whereas the present formalization of rationality-preserving weak-form PP and IP has theoretical appeal, in practical application it is vulnerable to contestability. 
Can there be widespread agreement within society about the appropriate values for $C_{FP}, C_{TN}, C_{FN}$, and $C_{TP}$? 
Can the same be said about the parameters of the sampling distributions, and the operative value for $\epsilon$?
In practical application, these parameters are heavily contested by opposing interest groups. 
The advantage of relaxing normative rationality to admit valuations based on moral imperatives or protected values is that the framework then delivers clear regulatory-policy recommendations \emph{within} each like-minded interest group. 
But at that point the PP and IP cease to be compatible guides for AI innovation governance. 

%

\subsection{winner takes all}
Underpinning the surge of investment in the development of AI foundational models is the view that the market for services powered by generative AI will be highly concentrated, dominated by the firm that can drive its technology ahead of the field and thereafter stay ahead. 
In other words, it has been viewed as a winner-takes-all market. 
The consequences of this arms-race are explored in numerous formal models. 
\citet{armstrong/etal:16-ais} study how teams in such technology-development races have incentives to cut corners and under-invest in safety precautions to increase the probability of winning the race. 
\citet{n/d:20-ais} develop an all-pay contest model for deriving public-policy measures to avoid the emergence of a poor-quality (possibly dangerous to humans) Artificial General Intelligence (AGI) out of the winner-takes-all race.\footnote{They recommend the taxing AI and using public procurement to reduce the probability of the emergence of an `unfriendly' AGI. The underlying logic being that these measures would reduce teams' payoffs, increase the total amount of R\&D investment needed, and incentivize co-operation and co-ordination. 
}
The paper argues for globally co-ordinated taxation and regulation of AI. 

To date, AI regulatory initiatives do not show clear evidence global co-ordination. 
In keeping with SDT-motivated limitations of regulatory sandboxes,\footnote{See discussion of the limitations of regulatory sandboxes on p \pageref{para:sandbox-limitations-from-SDT}.} there are no public calls by regulators requiring frontier AI system development to take place within regulatory sandboxes. 

Competition authorities are also concerned about the possibility of a `winner-takes-most' outcome among AI foundation models \citep{cma:23a}. 
At the international level, competition authorities have argued that competitive markets for AI services are needed to ensure that their benefits will be widely felt, and that ``if there are `winner-takes-all/most' characteristics, intervention risks being ineffective if too late'' \citep{oecd:24}. 

In many IT product markets, firms gain a dominant position through network effects and leveraging those network effects into neighboring markets --- i.e. through diffusion of their product through the economy and through their associated market conduct. 
In such IT product markets, capturing dominant positions and exploiting them takes place in the market-conduct space overseen by competition authorities. 
However the race for frontier AI-system dominance takes place in technology space, navigated by R\&D investment --- which is not conventionally the purview of competition authorities. 
Although it may seem that the IP and the related `pro-innovation principle' \citep{dsit:23,dsit:24} are ranged against national-level competition policy concerns, they in fact operate in different spaces --- the former in the technological space, the latter in the space of market conduct.

\section{CONCLUSION}\label{sec:conclusion}
The presently developed SDT-based model operationalizes the weak PP, and this informs fundamental questions pertaining to the general scope of the PP in relation to other decision rules. 
This model allows integrated consideration of factors and dimensions that previously have been considered in isolation. 

The first of these, which can be interpreted as the reason for the existence of the PP, is the bridging of the disparate inferential conventions and requirements of science (NHST) and policy decision making.
NHST inferential practices followed in science are intended to hold type-I error at the fixed, conventional level of $\alpha=.05$.
However policy decision making requires an inferential threshold that reflects operative cost tradeoffs.
The weak PP, understood as a verbal formulation of an  SDT corner solution, serves as a \emph{patch} for this disparity between the inferential conventions of science and the needs of practical policy decision making. 

The second is the non-separability of the `applicability of the PP' and the `nature of the intervention' questions. 
The third is the possibility to incorporate quasi-option value.
The fourth is the explicit incorporation of countervailing risks caused by intervention measures themselves. 
The fifth is the weak PP's relation to strong PP and super-strong PP, and the sixth is the natural complementarity between weak PP and weak IP as the two possible SDT corner solutions. 
Although there are other formalizations of the PP,\footnote{Recent examples include \citet{steel/bartha:22-ra} and \citet{boyer-kassem/duchene:24-jbe}.} the present work is the first to formalize (weak) IP and to do so in a single common framework together with (weak) PP. 

Furthermore, the SDT-based model provides a structure within which to analyze a variety of questions and issues arising in AI innovation governance. 
This paper focuses on sandboxing and how this regulatory instrument --- which is increasingly popular globally --- can be explained through the lens of SDT. 
Each key element of SDT --- discriminability, the expected misclassification cost ratio, and the qualitative properties of the optimal inferential threshold --- is reflected in the makeup and operation of regulatory sandboxes. 
In short, the SDT model provides decision-theoretic underpinning for regulatory sandboxing. 
This model suggests that AI foundation models are ill-suited for regulatory sandboxing, as their general-purpose nature entails that the set of applications by na\"{i}ve, well-behaved, and nefarious users is open-ended, with the consequence that the relevant misclassification costs cannot be uniquely identified.



\newpage

\newpage
\linespread{1.45}\small\normalsize
\appendix
\section*{APPENDICES}

\section{Scientific inferential convention: NHST}\label{app:NHST}
Null Hypothesis Significance Testing (NHST) is the workhorse method of statistical inference in modern science.
It combines the Neyman-Pearson concept of a critical rejection region \citep{neyman/pearson:33-ptrsla}  with Fisher's formulation of $p$-values \citep{fisher:59}.
Although there are basic, pointed philosophical differences between the developers of these two concepts,%
\footnote{The distinction between `inductive inference' as advocated by Fisher, and `inductive behavior' as advocated by Neyman, was at the heart of of their disagreement. Neyman advocated a theory of mathematical statistics predicated on probability (not subjective likelihood), the basis of which is provided by ``the conception of frequency of errors in judgement.'' \citep{neyman:35-jrss,lehmann:93-jasa} }
in modern usage these differences have been glossed over or subsumed within a unified framework \citep{lehmann:93-jasa,berger:03-ss}. 

That NHST has become a central preoccupation within empirical science was critically noted already by Yates \citep{yates:51-jasa}. 
Since Yates, criticism of this preoccupation and of NHST per se has been repeated and expanded \citep{nickerson:00-pm,sterne/smith:01-bmj,ziliak/mccloskey:08}. 
The widely cited paper entitled `Why most published research findings are false' by John Ioannidis  represents one culmination of this stream of criticism \citet{ioannidis:05-plosm}. 
Some of the strongest and most persistent critics of NHST are advocates of Bayesian statistical methodology \citep{kruschke:10-wire}. 
Nevertheless NHST remains the prevailing convention --- thus far, in all but one journal.%
\footnote{In 2015, the editors of \emph{Basic and Applied Social Psychology} announced that they will be removing $p$-values and other NHST measures from papers published in \emph{BASP} \citep{trafimow/marks:15-basp}. }

\subsection{Observations}\label{app:NHST-observations}
The present paper is not intended to augment the general critique of NHST.
Nevertheless I flag three observations which also feature in that literature.

First, note that the $\alpha=0.05$ level is, ostensibly, arbitrary \citep{sterne/smith:01-bmj,lehmann/romano:05}. 
Appendix \ref{sec:alpha-level} traces the broad outlines of how this convention arose, starting with the recommendations and statistical tables of Ronald Fisher.
In fact the $\alpha=0.05$ level is not a sufficiently demanding criterion that it would identify only strong evidence against the null.

Second, modern commentators such as David Cox are in agreement with Ronald Fisher, who held that drawing sharp distinctions between $p$-values such as 0.051 and 0.049 introduces an artificially sharp dichotomy \citep{cox:82-bjcp}. 
Ceteris paribus, the evidential value of a study supplying a $p$-value of 0.051 is virtually indistinguishable from that of a study supplying a $p$-value of 0.049.
Applying the labels `non-significant' to the former and `significant' to the latter facilitates dichotomous thinking --- where the underlying evidence does not in itself support such a distinction.

Third, `statistical significance' is not synonymous with `scientific significance' \citep{cox:82-bjcp}. 
The connection with policy-making relevance is even more tenuous.
For instance observational studies can achieve statistical significance by virtue of sample size, but the effect size may be minuscule, contributing little to overall scientific understanding or to the understanding of effective policy levers for decision making.

However, as is shown in Section \ref{sec:classical-sdt}, these three detractions lose force when a fixed $\alpha$ is abandoned in favor of a contextually optimal inferential threshold $\alpha^*$.

\section{The fixed \texorpdfstring{$\boldsymbol{\alpha=0.05}$}{0.05-level} threshold}\label{sec:alpha-level}

Fisher introduced significance testing and the concept of a $p$-value, i.e.\ the probability that a test-statistic $T=t(X)$,\footnote{computed on observed data drawn from a continuous distribution $X \sim f(x|\theta)$ on support $\mathbb{R}$} equals or exceeds the observed value $t(x)$ given that the null hypothesis $H_0: \theta=\theta_0$ is true, i.e.\ $p=P(t(X)\geq t(x)|H_0)$.
In Fisher's approach to significance testing, there is no explicit alternative hypothesis under consideration.
This is because there are innumerable different conceivable alternative hypotheses.
Fisher views the alternative hypothesis --- and therefore any quantities derived from it, such as statistical power --- as `unknown'.
Although Fisher believed that $p$-values require researchers' \emph{subjective} interpretation, his early expositions advocated using $p<0.05$ (i.e.\ a 5\% significance level) as the standard for concluding that there is evidence against $H_0$.

\begin{quote}
\citet{fisher:25}: The value for which P $=$ .05, or 1 in 20, is 1.96 or nearly 2; it is convenient to take this point as a limit in judging whether a deviation is to be considered significant or not. ... ...We shall not often be astray if we draw a conventional line at 0.05 ... .

\citet{fisher:26-jmagb}: Personally, the writer prefers to set a low standard of significance at the 5 percent point, and ignore entirely all results which fail to reach this level.

\citet{fisher:35}: It is usual and convenient for experimenters to take 5 percent as a standard level of significance, in the sense that they are prepared to ignore all results which fail to reach this standard... .
\end{quote}

\noindent Fisher viewed the $p$-value as an index of the `strength of evidence' against $H_0$.
Fisher's approach to significance testing thus focuses on controlling type-I error alone.
Although in his later work Fisher attacked the notion of a standard or conventional threshold for type-I error, empirical researchers continue to employ the $\alpha = 0.05$ level suggested by Fisher.
Fisher's influential texts included tabulations of exact small-sample $\mathcal{X}^2$-,\: $t$- and $F$-test statistics.
He economized on page-space and enhanced the usability of his tables by providing only selected quantiles, key among which being the 5\% quantile.
Neyman and Pearson followed suit in endorsing a fixed 5\% level --- and in turning their attention to controlling type-I error and in developing their method around a `rule of behavior' --- under the influence of Fisher's 5\% and 1\% quantile tables \citep{lehmann:93-jasa}.

Neyman and Pearson held that one could only test a null hypothesis \emph{against} an alternative hypothesis.
Thus Neyman and Pearson were concerned with type-II error as well as type-I error.
Following this concern, they introduced the concept of statistical power.
They sought to supplant the subjective element present in Fisher's approach with a formalized decision procedure (a behavioral rule) embodying the frequentist principle:
``In repeated practical use of a statistical procedure, the long-run average actual error should not be greater than (and ideally should equal) the long-run average reported error'' \citep{berger:03-ss}.
Neyman and Pearson sought to distinguish their theory from Fisher's `significance testing', and did so by referring to their formalized decision rule as `hypothesis testing'.

\begin{statement}[Neyman-Pearson hypothesis testing]\label{st:npl}~\par
\begin{itemize}
\item[(i)] Derive type-I and type-II error probabilities $\alpha = P(t(X) \geq c\,|H_0)$ and $\beta=P(t(X) < c\,|H_1)$ for given for simple hypotheses $H_0:\theta=\theta_0$ and $H_1:\theta=\theta_1$ where $X \sim f(x|\theta_i),\; i=\{0,1\}$,\; $\theta_1 > \theta_0$, and $c$ is a critical threshold in the codomain of $t(\cdot)$;
\item[(ii)] Determine the most powerful test (in particular its critical threshold $c$) and the most appropriate type-I error probability\; $\alpha^*$\; using $\alpha=P(t(X) \geq c\,|H_0)$, $\beta=P(t(X) < c\,|H_1)$, $X \sim f(x|\theta_i)$, and the costs associated with type-I and type-II errors;
\item[(iii)] Use the pre-chosen critical value $c$ to reject $H_0$ if $t(X) \geq c$, else accept $H_0$.
\end{itemize}
\end{statement}

Notice that there are two components in Part (ii) of this statement.
The first is the determination of the most powerful test.
This is accomplished with the Neyman-Pearson lemma.
The second is the determination of the most appropriate type-I error probability $\alpha^*$.
For this, Neyman and Pearson did not provide a formal procedure, but offered clear verbal guidance.
I elaborate the Neyman-Pearson lemma first, followed by $\alpha^*$, even though the latter is technically a required input parameter for application of the Neyman-Pearson lemma.
The following presentation of the Neyman-Pearson lemma is adapted from \citep{lehmann/romano:05}, which may also be consulted for the associated proof.

\begin{theorem}[Neyman-Pearson lemma] Let there be two continuous distributions $X \sim f(x|\theta_i),\; i=\{0,1\}$,\; indexed by the parameters $\theta_1 > \theta_0$.
\begin{itemize}
\item[(i)] Existence. For testing the simple null hypothesis $H_0:\theta=\theta_0$ against the simple alternative hypothesis $H_1:\theta=\theta_1$,  there exists a test function $\phi$ and a constant $k>0$ such that
    \begin{equation} \label{eq:level}
    E_{\theta_0}\phi(X)=\alpha
    \end{equation}
    and
    \begin{equation} \label{eq:likelr}
    \phi(x)=
    \begin{cases}
    1\;\;\; \mathrm{if}\;\;\;\frac{f(x|\theta_1)}{f(x|\theta_0)}\: >\: k \\
    0\;\;\; \mathrm{if}\;\;\;\frac{f(x|\theta_1)}{f(x|\theta_0)}\: <\: k
    \end{cases}
    \end{equation}
\item[(ii)] Sufficient condition for a most powerful test. If $\phi$ satisfies \eqref{eq:level} and \eqref{eq:likelr} for some constant $k$, then $\phi$ is Most Powerful (MP) for testing $H_0$ against $H_1$ at level $\alpha$.
\item[(iii)] Necessary condition for a most powerful test. If a test $\phi^*$ is MP at level $\alpha$, then it satisfies \eqref{eq:likelr} for some $k$, and it also satisfies \eqref{eq:level} unless there exists a test of size strictly less than $\alpha$ with power 1.
\end{itemize}
\end{theorem}

Although the Neyman-Pearson lemma is framed in terms of simple hypotheses, the test $\phi^*$ can be shown to be Uniformly MP against a composite alternative hypothesis when the family of distributions indexed by $\theta_i$ satisfies the monotone likelihood ratio property.

Neyman and Pearson explicitly acknowledge that the critical threshold $c$, which demarcates between the null-hypothesis rejection region and the null-hypothesis acceptance region, should be determined by the researcher.
This determination is dependent upon the context:
\begin{quote}
...in some cases it will be more important to avoid the first [type-I error], in other the second [type-II error]... ...determining just how the balance should be struck, must be left to the investigator. 
... ...we attempt to adjust the balance between the risks [of the two types of error] to meet the type of problem before us. 
\citep{neyman/pearson:33-ptrsla}
\end{quote}
In this 1933 formulation, consideration of consequences --- costs of error --- remain implicit.
With time Neyman's position shifted, however.
In 1950 he articulated the view that controlling type-I errors is `more important' than controlling type-II errors:
\begin{quote}
Because an error of the first kind is more important to avoid than an error of the second kind, our requirement is that the test should reject the hypothesis tested when it is true very infrequently... ...The ordinary procedure is to fix arbitrarily a small number $\alpha$... ...and to require that the probability of committing an error of the first kind does not exceed $\alpha$.
\citep{neyman:50}
\end{quote}

From these beginnings, inertia took hold \citep{cowles/davis:82-ap}.
Today, use of $\alpha=0.05$ reflects a customary, conventional, common frame of reference:
\begin{quote}
It is customary therefore to assign a bound to the probability of incorrectly rejecting [$H_0$] when it is true and to attempt to minimize the other probability subject to this condition.
... ...The choice of a level of significance $\alpha$ is usually somewhat arbitrary... ...Standard values, such as .01 or .05, were originally chosen to effect a reduction in the tables needed for carrying  out various test [sic]. By habit, and because of the convenience of standardization in providing a common frame of reference, these values gradually became entrenched as the conventional levels to use. 
\citep{lehmann/romano:05}
\end{quote}

The key feature of operating under the Neyman-Pearson lemma is accepting --- as given, short of sample-size considerations --- the maximum achievable statistical power\; $1-\beta = E_{\theta_1}\phi(X)$\; associated with level $\alpha$.
This is equivalent to fixing $\alpha$ on the abscissa of the Receiver Operating Characteristics (ROC) space,  and accepting as given the associated \emph{power} as indicated by the ordinate of the ROC curve, i.e.\ the locus of all $(\alpha,1-\beta)$ points obtained parametrically by varying the cutoff threshold, given the distributions $X \sim f(x|\theta_i),\; i=\{0,1\}$.
Neither NHST nor the Neyman-Pearson lemma supports any explicit consideration of \emph{tradeoffs} between type-I and type-II errors.

\section{Derivations supporting Condition \ref{con:epsilon1-tolerance}}\label{sec:appendix-epsilon1}
Let $\epsilon$ be the maximum probability of inferential surprise that the regulator is willing to tolerate in issuing a weak-PP red-light determination or a weak-IP green-light determination. 
In other words, given the current discriminating ability of scientific experiment and analysis $d^\prime$, the regulator is willing to issue a weak-PP red-light determination if the optimal threshold $x^\prime$ is such that the associated probability of surprise is bounded by $\epsilon$:  
\begin{align}
    1-p(x^\prime < x) &\leq \epsilon\\
    1-[ P(\theta_0)\alpha^\prime + (1-P(\theta_0))(1-\beta^\prime)]  &\leq \epsilon\\
     P(\theta_0)(1-\alpha^\prime) + (1-P(\theta_0))\beta^\prime  &\leq \epsilon \label{ineq:a3}
\end{align}
The ROC curve does not fall below the principal diagonal for all $d^\prime\geq 0$, and thus 
\begin{equation}\label{ineq:a4}
    1-\beta \geq \alpha\qquad \Leftrightarrow\qquad 1-\alpha \geq \beta\;\;.
\end{equation}
Using this, notice that inequality \eqref{ineq:a3} continues to be satisfied if we substitute $1-\alpha^\prime$ with $\beta^\prime$. 
But instead substituting $\beta^\prime$ with $1-\alpha^\prime$ we define the upper bound of inferential surprise $\epsilon_1$\; ($\epsilon\leq \epsilon_1$) as 
\begin{align}
     P(\theta_0)(1-\alpha^\prime) + (1-P(\theta_0))(1-\alpha^\prime)  &\leq \epsilon_1\\
     (1-\alpha^\prime)  &\leq \epsilon_1 \label{ineq:a6}
\end{align}
Let $\alpha^\prime_{\epsilon_1}$ be the value that solves \eqref{ineq:a6} as an equality:
\begin{equation}
    (1-\alpha^\prime_{\epsilon_1}) = \epsilon_1\;\;. \label{ineq:a6-as-equality}
\end{equation}
Turning to the $\delta$-neighborhood of (1,1) 
\begin{align}
\sqrt{(1-\alpha^\prime)^2 + (1-(1-\beta^\prime))^2} &\leq \delta \\
      \sqrt{(1-\alpha^\prime)^2 + (\beta^\prime)^2} &\leq \delta
\end{align}
Making use of \eqref{ineq:a4} again and defining $\delta_1$\; ($\delta \leq \delta_1$) to be the upper bound of the neighborhood-defining radius
\begin{align}
      \sqrt{(1-\alpha^\prime)^2 + (1-\alpha^\prime)^2} &\leq \delta_1 \\
                      \sqrt{2(1-\alpha^\prime)^2} &\leq \delta_1 \\
                        \sqrt{2}(1-\alpha^\prime) &\leq \delta_1 \label{ineq:a11}
\end{align}
Solving for upper bound $\delta_1$ that corresponds with $\epsilon_1$ in \eqref{ineq:a6-as-equality} gives
\begin{align}
    \sqrt{2} (1-\alpha^\prime_{\epsilon_1}) &= \delta_1\\
    \sqrt{2}\cdot \epsilon_1 &=\delta_1
\end{align}

\section{Derivations supporting Condition \ref{con:epsilon0-tolerance}}\label{sec:appendix-epsilon0}
Let $\epsilon$ be the maximum probability of inferential surprise that the regulator is willing to tolerate in issuing a weak-PP red-light determination or a weak-IP green-light determination. 
In other words, given the current discriminating ability of scientific experiment and analysis $d^\prime$, the regulator is willing to issue a weak-IP green-light determination if the optimal threshold $x^\prime$ is such that the associated probability of surprise is bounded by $\epsilon$
\begin{align}
     1-p(x < x^\prime) &\leq \epsilon\\
     1 - [P(\theta_0)(1-\alpha^\prime) + (1-P(\theta_0))\beta^\prime]  &\leq \epsilon\\
     P(\theta_0)\alpha^\prime + (1-P(\theta_0))(1-\beta^\prime)  &\leq \epsilon \label{ineq:b3}
\end{align}
Using \eqref{ineq:a4}, notice that inequality \eqref{ineq:b3} continues to be satisfied if we substitute $1-\beta^\prime$ with $\alpha^\prime$. 
But instead substituting $\alpha^\prime$ with $1-\beta^\prime$ we define the upper bound of inferential surprise $\epsilon_0$\; ($\epsilon\leq \epsilon_0$) as 
\begin{align}
     P(\theta_0)(1-\beta^\prime) + (1-P(\theta_0))(1-\beta^\prime)  &\leq \epsilon_0\\
                                             (1-\beta^\prime)  &\leq \epsilon_0 \label{ineq:b5}
\end{align}
Let $\beta^\prime_{\epsilon_0}$ be the value that solves \eqref{ineq:b5} as an equality:
\begin{equation}
    (1-\beta^\prime_{\epsilon_0}) = \epsilon_0\;\;. \label{ineq:b5-as-equality}
\end{equation}
Let the $\delta$-neighborhood of (0,0) be
\begin{equation}
    \sqrt{(\alpha^\prime)^2 + (1-\beta^\prime)^2} \leq \delta    
\end{equation}
Making use of \eqref{ineq:a4} again and defining $\delta_0$\; ($\delta \leq \delta_0$) to be the upper bound of the neighborhood-defining radius, then 
\begin{align}
    \sqrt{(1-\beta^\prime)^2 + (1-\beta^\prime)^2} &\leq \delta_0 \\
    \sqrt{2(1-\beta^\prime)^2} &\leq \delta_0 \\
    \sqrt{2}(1-\beta^\prime) &\leq \delta_0 \label{ineq:b9}
\end{align}
Solving for the upper bound $\delta_0$ that corresponds with $\epsilon_0$ in \eqref{ineq:b5-as-equality} gives
\begin{align}
    \sqrt{2}(1-\beta^\prime_{\epsilon_0}) &= \delta_0 \\
    \sqrt{2} \cdot \epsilon_0 &= \delta_0
\end{align}


\end{document}